\documentclass[fleqn,usenatbib]{mnras}

\usepackage{newtxtext,newtxmath}

\usepackage[T1]{fontenc}

\DeclareRobustCommand{\VAN}[3]{#2}
\let\VANthebibliography\thebibliography
\def\thebibliography{\DeclareRobustCommand{\VAN}[3]{##3}\VANthebibliography}

\usepackage{graphicx}	
\usepackage{amsmath}	
\usepackage{deluxetable}
\usepackage{adjustbox}
\usepackage[capitalise]{cleveref}

\newcommand\gleamxfull{GLEAM-X\,J162759.5--523504.3}
\newcommand\gleamx{GLEAM-X\,J1627--52}
\newcommand{\ha}{H$\alpha$}
\newcommand{\hb}{H$\beta$}
\newcommand{\kms}{\,km\,s$^{-1}$}

\title[The counterpart of \gleamx]{Constraints on optical and near-infrared variability in the localisation of the long-period radio transient \gleamx}

\author[Lyman et al.]{
J. D. Lyman,$^{1}$\thanks{E-mail: J.D.Lyman@warwick.ac.uk (JDL)}
V. S. Dhillon,$^{2,3}$
S. Kamann,$^{4}$
A. A. Chrimes,$^{5,6}$
A. J. Levan,$^{6,1}$
I. Pelisoli,$^{1}$
\newauthor
D. T. H. Steeghs$^{1}$
and K. Wiersema$^{7}$
\\
$^{1}$Department of Physics, University of Warwick, Coventry, CV4 7AL, UK\\
$^{2}$Astrophysics Research Cluster, School of Mathematical and Physical Sciences, University of Sheffield, Sheffield S3 7RH, UK\\
$^{3}$Instituto de Astrof\'{i}sica de Canarias, E-38205 La Laguna, Tenerife, Spain\\
$^{4}$Astrophysics Research Institute, Liverpool John Moores University, IC2 Liverpool Science Park, 146 Brownlow Hill, Liverpool L3 5RF, UK\\
$^{5}$European Space Agency (ESA), European Space Research and Technology Centre (ESTEC), Keplerlaan 1, 2201 AZ Noordwijk, the Netherlands\\
$^{6}$Department of Astrophysics/IMAPP, Radboud University Nijmegen, P.O. Box 9010, 6500 GL Nijmegen, The Netherlands\\
$^{7}$Centre for Astrophysics Research, University of Hertfordshire, Hatfield AL10 9AB, UK\\
}

\date{Accepted XXX. Received YYY; in original form ZZZ}

\pubyear{\the\year{}}

\begin{document}
\label{firstpage}
\pagerange{\pageref{firstpage}--\pageref{lastpage}}
\maketitle

\begin{abstract}
\gleamx\ was discovered as a periodic ($\sim$18\,min) radio signal over a duration of three months in 2018. It is an enigmatic example of a growing population of `long-period radio transients' consistent with Galactic origins. Their nature is uncertain, and leading models invoke magnetic neutron stars or white dwarfs, potentially in close binary systems, to power them. \gleamx\ resides in the Galactic plane with a comparatively coarse localisation ($\simeq$2\,arcsecond). Here we study the localisation region to search for spectrophotometric signatures of a counterpart using time-domain searches in optical and near-infra-red imaging, and MUSE integral field spectroscopy. No sources in the localisation display clear white dwarf spectral signatures, although at the expected distance we can only provide modest limits on their presence directly. We rule out the presence of hot sub-dwarfs in the vicinity. We found no candidate within our search for variability or periodic behaviour in the light curves. Radial velocity curves additionally show only weak evidence of variation, requiring any realistic underlying system to have very low orbital inclination ($i \lesssim 5$\,deg). Two Balmer emission line sources are reminiscent of white dwarf pulsar systems, but their characteristics fall within expected M-dwarf chromospheric activity with no signs of being in a close binary. Currently the white dwarf pulsar scenario is not supported, although longer baseline data and data contemporaneous with a radio active epoch are required before stronger statements. Isolated magnetars, or compact binaries remain viable. Our limits highlight the difficulty of these searches in dense environments at the limits of ground-based data.

\end{abstract}




\section{Introduction}
\label{sec:intro}

Cadenced wide-field radio surveys at both high- and low-frequencies are beginning to probe the variable and transient radio sky over increasingly wide ranges of timescales and luminosities. Well-established Galactic populations of pulsars \citep{manchester2005} and magnetars \citep{olausenkaspi2014} have been known for many years and are seen to display periodicity of emission on timescales of $\lesssim$10 seconds. Conversely, a small subset of the more-recently discovered extra-galactic fast radio bursts (FRBs) show repeating bursts on timescales of $\sim0.4-40$\,days \citep[][although notable exceptions have displayed bursts at higher rates for periods of time, e.g. \citealt{lanman2022}]{chime2023}. Rotating (and magnetised) neutron stars naturally explain these Galactic populations \citep[e.g.][]{lyne1985, thompsonduncan1995} and are also a leading model for repeating FRBs \citep{cordes2019}. Between these two time-scales, an emerging population of long-period Galactic radio transients is been revealed \citep[e.g.][]{hyman2005, hurleywalker2023, caleb2022, hurleywalker2024, deruiter2024, dong2024, li2024, lee25} that is challenging current models. The population is grouped on displaying periodic coherent radio emission on minutes to hour timescales, however they are otherwise relatively heterogeneous. Some sources display regular period emission over long timescales (years), with others seemingly active only for short periods and/or showing only intermittent burst emission. The pulse profiles of the emission are also characteristically diverse, even for a given source. Rotating Radio Transients \citep{mclaughlin2006} share some of these characteristics, albeit without clear periodicity in the burst emission, a defining characteristic of these long-period radio transients.

Among the emerging population of long-period Galactic radio transients is \gleamxfull\ (hereafter \gleamx). Reported by \citet{hurleywalker2022}, \gleamx\ is a Galactic Plane source at a distance of $1.3\pm0.4$\,kpc, displaying heterogeneous pulse profiles with a period \mbox{$P = 1091.169$\,seconds}, which was seemingly active for only 3 months in early 2018. 
A strong linear polarisation of the pulses was used to infer the presence of strong magnetic fields, with rapid variability indicating a compact emitting region. Taken together, \citet{hurleywalker2022} use these arguments to favour a magnetar origin for the emission. 

The rotation rate of magnetars slows over time due to energy losses, mainly magnetic dipole emission \citep{pacini1967} and a measure of the current spin period and its derivative under this model can be used to infer the age and magnetic field strength of the magnetar \citep{duncanthompson1992}. Additionally, as the period of pulsar and magnetar emission is determined by their spin, $\sim$minutes--hour long periodicity of emission would typically be way beyond the pulsar ``death line" \citep{chen1993} where emission is expected to cease, except for extreme magnetic field strengths. Following this, the properties of \gleamx\ suggest it may have $B \gtrsim 10^{16}$\,G, and consequently the most magnetised neutron star known \citep{suvorov2023, konar2023}. \gleamx\ is therefore of chief interest for understanding the evolution of compact objects as the final stages of stellar evolution, with implications for other transient populations \citep{beniamini2023}.
It is possible to produce such long period systems in the magnetar scenario via fallback disk braking \citep{chatterjee2000}. Indeed, the fall-back accretion scenario as a braking mechanism is the favoured model to explain the slowest rotational period magnetar, with a period of 6.7\,hours \citep{rea2016}. As noted by \citet{tong2023}, \gleamx\ may then represent an intermediate object in a poorly-observed regime between the extrema of magnetar periods observed. Using population synthesis, \citet{rea2024} recently conclude that isolated neutron stars, emitting as the classical rotating dipole pulsar, are an unlikely scenario to explain most, if not all, long-period radio transients.

Other analyses have favoured a highly-magnetised White Dwarf (WD) or hot sub-dwarf origin for \gleamx\ \citep[e.g.][]{katz2022, loebmaoz2022} wherein the large moment of inertia of the stellar system (cf. that of a neutron star) provides a sufficient rotational energy reservoir to power the observed radio emission. However, as noted by \citet{konar2023}, the required magnetic field strengths remain significantly larger than those seen in typical WD or hot sub-dwarf systems. Periodic radio emission has been seen in three WD systems. The AR Sco \citep{marsh2016} and J$1912-4410$ \citep{pelisoli2023} systems display dominant periodic emission on minutes time-scale (associated with the WD spin), with hours-long modulation on the timescale of their orbits with low-mass non-degenerate companions, and form a class of `WD pulsar'. Conversely, the recently discovered ILT\,J1101+5521 system \citep{deruiter2024} displays periodic emission on a 125.5\,minute timescale that matches the orbital period of the system. In the former systems, the leading explanations for the pulsed radio emission require interaction between the WD and its non-degenerate companion -- either through magnetic reconnection between the fields of the WD and the companion producing synchrotron emission \citep[e.g][]{Katz2017,garnavich2019}, or through electron-cyclotron maser when particles seeded by the companion reach the white dwarf polar region \citep{pelisoli2024}. In ILT\,J1101+5521, the origin is even less clear, although the binary is likely to be in a polar configuration with a moderately magnetic WD, which could produce radio pulses due to geometry effects of a beam sweeping with the orbit. 

As with many astrophysical phenomena, unlocking the true nature of long-period radio transients may be largely contingent on multi-wavelength characterisation. This will allow for the nature of their underlying systems and powering mechanisms to be uncovered, and ultimately place them in the landscape of stellar and compact object evolution. Unfortunately, observationally and/or intrinsically rare Galactic phenomena are typically found in areas of high stellar density and line-of-sight extinction, hampering optical and  near-infrared (NIR) counterpart searches.

GPM J1839–10 \citep{hurleywalker2023}, which displays $\sim$minute long radio bursts on a 21\,min period and has been active for at least 30 years, has a promising candidate K/M-dwarf NIR counterpart identified. The large distance to this object (5.7$\pm$2.9\,kpc) gives confidence that such counterparts may be found for other nearby events. As alluded to above, a definitive counterpart binary system has been associated to the ILT\,J1101+5521 system \citep{deruiter2024}, harbouring an M-dwarf and WD in an orbit matching the 2.1\,hour periodicity of the radio emission. Most recently, the WD interpretation for minutes-long period transients has received support thanks to the recent detection in \citet{hurleywalker2024} of an optical counterpart for GLEAM-X\,J0704--37. The source is an M-dwarf with a very low chance of spatial coincidence thanks to the unusually high Galactic latitude of this system. \citet{hurleywalker2024} note that a stellar origin from a lone M-dwarf is unlikely, and they disfavour a M-dwarf -- neutron star system, preferring a M-dwarf -- WD binary. The binary nature of this system has since been conclusively identified by \citet{rodriguez2025}, who found an orbital period matching that of the radio period ($2.9$\,hr), joining ILT\,J1101+5521 in this regard. The spectra reveal a comparatively massive WD companion to the M-dwarf in GLEAM-X\,J0704--37 at 0.8--1.0\,M$_\odot$. Interestingly, renewed calibration of the source flux also significantly reduces the inferred distance of this source at $\sim400$\,pc \citep{rodriguez2025}, in comparison to the original estimate of $\sim1.5$\,kpc \citep{hurleywalker2024}.

For \gleamx, deep X-ray observations have been able to place constraints on high-energy emission from the system \citep{rea2022}, but the challenge of detecting a multi-wavelength counterpart in the optical or NIR in its crowded location is compounded by a comparatively crude localisation from the radio emission. \citet{rea2022} obtained spectroscopy for three of the brighter candidate sources in the localisation, finding them to be typical F, G, K stars, with no remarkable features.

Here we present a comprehensive search for optical and NIR spectrophotometric variability in the vicinity of \gleamx\ as a means to identify candidate counterparts. Models detailed above place requirements on the nature of the optical counterparts stellar type and/or its existance in a tight binary configuration. Our data will be used to probe for the presence of peculiar stellar types, photometric variability, and radial velocity curves in sources local to \gleamx, as a means to identify a candidate counterpart. Where appropriate, we assume a distance to the source of $1.3\pm0.5$\,kpc based on its radio dispersion measure \citep{rea2022}. In \cref{sec:observations} we present our observations of the field of \gleamx. We present our methods to search for an optical counterpart in \cref{sec:methods} and our results in \cref{sec:results}. These are discussed in \cref{sec:discussion}.

\section{Observations and Data reduction}
\label{sec:observations}

In order to probe any optical counterpart of \gleamx, observations were acquired from three facilities of the European Southern Observatory (ESO) in Chile: the Multi Unit Spectroscopic Explorer (MUSE) mounted on UT4 of the 8.2m Very Large Telescope (VLT), ULTRACAM mounted on the 3.5m New Technology Telescope, and survey data from the 4.1m Visible and Infrared Survey Telescope for Astronomy (VISTA). Each of these data sets is described separately in the following sections.

\subsection{VLT/MUSE}
\label{sec:obs_vltmuse}

The MUSE instrument \citep{bacon10} is an integral-field unit (IFU) offering seeing-limited spatially resolved spectroscopy over a $\sim$1\,arcmin field of view (FoV). This FoV comfortably covers the localisation uncertainty ($\sim$2\,arcsecond) of \gleamx, and provides optical spectroscopy from $4800$--$9300$\,\r{A} ($R\sim1800$--$3600$) across the field and so of every detected source. Observations were taken over three separate Observing Blocks (OBs) during late May 2022. Each OB consisted of $7\times313$\,second exposures with derotator offsets of 45\,degree between subsequent exposures. The final exposure for the first OB was cut short to 62.8\,seconds due to the approach of morning twilight. Details of these epochs are shown in \cref{tab:muse_epochs}.

Natural seeing was improved upon during observations through the use of a local Tip-Tilt star with the Wide-Field Mode Adaptive Optics (AO) of MUSE. Although this AO mode causes a region of unusable data in the spectral region surrounding the \ion{Na}{1}\,D doublet, this did not compromise any of the science goals of these data. No separate sky observations were obtained -- instead spaxels void of astrophysical signal within the FoV were used for sky subtraction and calibration. Individual exposures within the OBs were reduced and then combined with the MUSE data reduction pipeline \citep[][version 2.8.5]{weilbacher2020} within the {\sc EsoReflex} environment from ESO \citep{esoreflex13}. Sky-residuals were further removed using the Zurich Atmospheric Purge (ZAP; \citealp{zap16}) software version 2.1 with default parameters. For input into ZAP we constructed an aggressive source mask by calculating a sigma-clipped median and standard deviation, and flagging all pixels greater than one standard deviation above the median value. The 20 full-length exposures from all OBs (i.e. excluding the shortened final exposure of OB1, epoch 7) were also combined to generate a deep-stacked cube. In total therefore we have data-cubes covering 25 ``epochs'' of MUSE observations: 21 numbered from the individual exposures, OB1, OB2, OB3 from the stacked OBs and 1 `all' from the total exposure stack. Both individual exposures and the stacked OBs had FWHM measurements of 0.6--0.7'', as measured from bright stars in the field.

\begin{table}
\caption{Observing blocks of MUSE data covering \gleamx\ as referred to in the text. Stacked epochs are composed as follows: OB1 [1-7], OB2 [8-14], OB3 [15-21], all [1-6, 8-21]}
\label{tab:muse_epochs}
\begin{tabular}{cccc}
\hline
Epoch Name & Start time & End time & Exptime \\
           & (UTC)      & (UTC)    & (seconds) \\
\hline
1     & 2022-05-25 06:31:32 & 2022-05-25 06:36:45 & 313.0  \\
2     & 2022-05-25 06:38:47 & 2022-05-25 06:44:00 & 313.0  \\
3     & 2022-05-25 06:46:02 & 2022-05-25 06:51:15 & 313.0  \\
4     & 2022-05-25 06:53:16 & 2022-05-25 06:58:29 & 313.0  \\
5     & 2022-05-25 07:00:29 & 2022-05-25 07:05:42 & 313.0  \\
6     & 2022-05-25 07:07:44 & 2022-05-25 07:12:57 & 313.0  \\
7     & 2022-05-25 07:15:00 & 2022-05-25 07:16:02 & 62.8   \\
OB1   & 2022-05-25 06:31:32 & 2022-05-25 07:16:02 & 1940.8 \\
8     & 2022-05-28 01:11:20 & 2022-05-28 01:16:33 & 313.0  \\
9     & 2022-05-28 01:18:37 & 2022-05-28 01:23:50 & 313.0  \\
10    & 2022-05-28 01:26:53 & 2022-05-28 01:32:06 & 313.0  \\
11    & 2022-05-28 01:34:10 & 2022-05-28 01:39:23 & 313.0  \\
12    & 2022-05-28 01:41:27 & 2022-05-28 01:46:40 & 313.0  \\
13    & 2022-05-28 01:48:44 & 2022-05-28 01:53:57 & 313.0  \\
14    & 2022-05-28 01:56:00 & 2022-05-28 02:01:13 & 313.0  \\
OB2   & 2022-05-28 01:11:20 & 2022-05-28 02:01:13 & 2191.0 \\
15    & 2022-05-28 02:06:40 & 2022-05-28 02:11:53 & 313.0  \\
16    & 2022-05-28 02:13:54 & 2022-05-28 02:19:07 & 313.0  \\
17    & 2022-05-28 02:21:12 & 2022-05-28 02:26:25 & 313.0  \\
18    & 2022-05-28 02:28:27 & 2022-05-28 02:33:40 & 313.0  \\
19    & 2022-05-28 02:35:40 & 2022-05-28 02:40:53 & 313.0  \\
20    & 2022-05-28 02:42:56 & 2022-05-28 02:48:09 & 313.0  \\
21    & 2022-05-28 02:50:12 & 2022-05-28 02:55:25 & 313.0  \\
OB3   & 2022-05-28 02:06:40 & 2022-05-28 02:55:25 & 2191.0 \\
all   & 2022-05-25 06:31:32 & 2022-05-28 02:55:25 & 6260.0 \\
\hline
\end{tabular}
\end{table}

Spectra of sources within each data-cube were extracted with the {\sc pampelmuse} software \citep{pampelmuse2013, pampelmuse2018}. In the absence of a higher-resolution image from which to build a source catalogue, as is typically done with {\sc pampelmuse}, and, given our field is not overly crowded, we instead built source catalogues directly from each MUSE epoch using a DAOPhot \citep{stetson1987} style star-finding algorithm to detect any source with a peak more than four times the background rms in white-light images formed from the cubes. {\sc pampelmuse} was then run using the recommended procedure and parameter choices\footnote{Using documentation contained within \url{https://gitlab.gwdg.de/skamann/pampelmuse/}.} to extract spectra at the location of each of these detections. The extraction process takes care of the changing position and shape of the point-spread function throughout the spectral-axis of the MUSE data cubes. As our initial source catalogue contains even marginal detections in the stacked cube, some sources could not have useful spectra extracted. The very low signal-to-noise ratio (SNR) of faint sources in short wavelength ranges of the cube precluded the calculation of wavelength-dependent centroid and point-spread function evolution by {\sc pampelmuse}. In practice, such marginal sources in our stacked cube would not allow for useful insights into their origin even if spectra were extracted.

An absolute WCS solution is not performed during ESO data reduction of MUSE data and is instead populated by the telescope pointing coordinates which may be inaccurate by a few arcseconds. Therefore, to accurately cross-match sources detected in individual epoch catalogues we tied each data-cube's WCS to that of a Dark Energy Camera (DECam; \citealt{flaugher2015}) image accessed via the NOIRLab archive.\footnote{\url{https://astroarchive.noirlab.edu/}} The alignments were done using {\sc spalipy} \citep{spalipy} to calculate and perform source-based affine transformation of the MUSE data onto the pixel space of the DECam image. Alignment residuals were $\sim0.4$--$0.5$ MUSE pixels ($\sim$0.1\,arcsec). This is much smaller than our source density and allowed unambiguous source matching between catalogues. This also allowed us to place the sources on an accurate absolute WCS plane for locating \gleamx\ and searching within its localisation uncertainty. Due to varying data quality, each source was not able to be extracted from every epoch. Each unique source was assigned an arbitrary id based on ascending declination, which will be referred to in later sections.

\begin{figure*}
\includegraphics[width=0.49\linewidth]{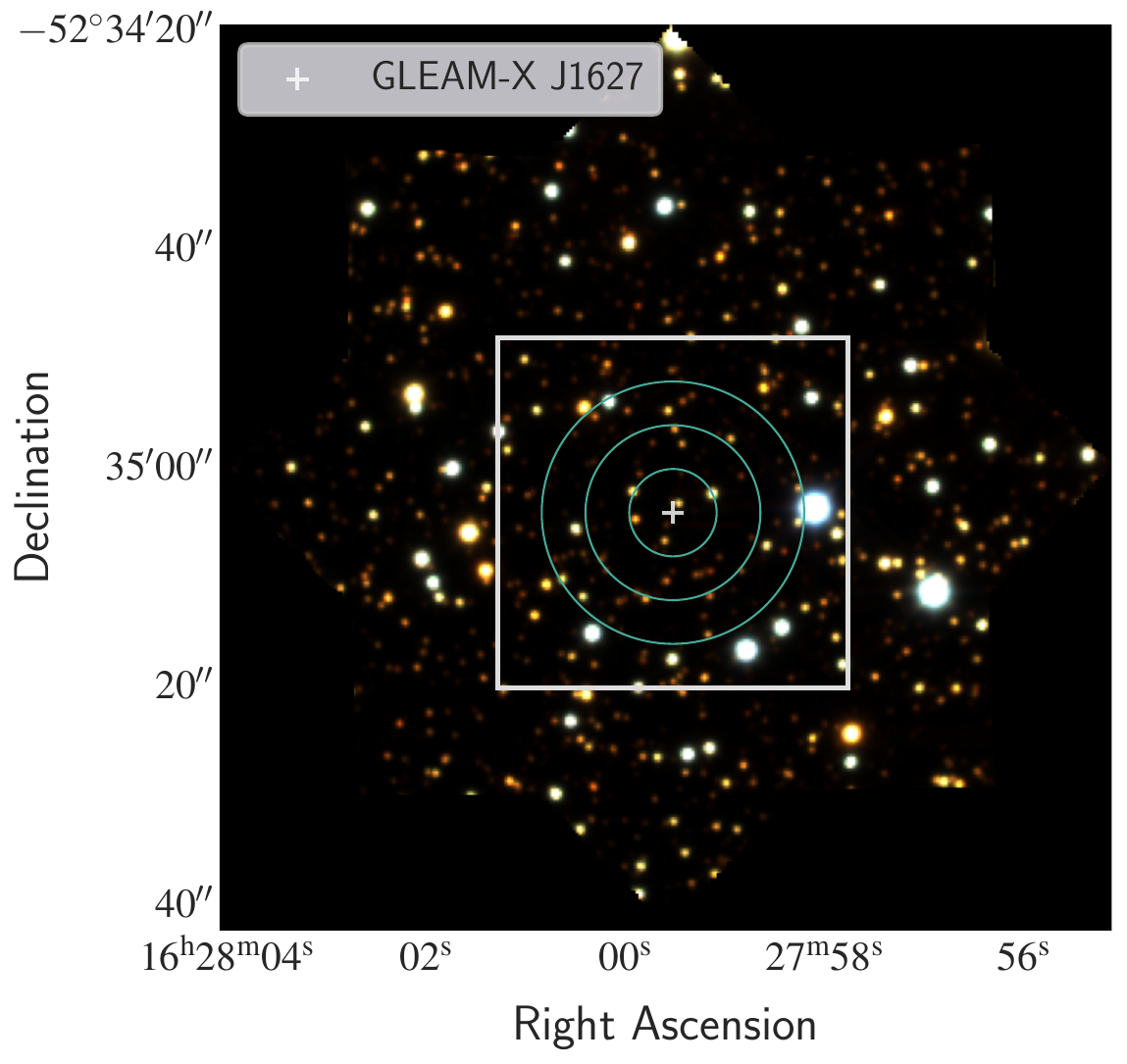}
\includegraphics[width=0.49\linewidth]{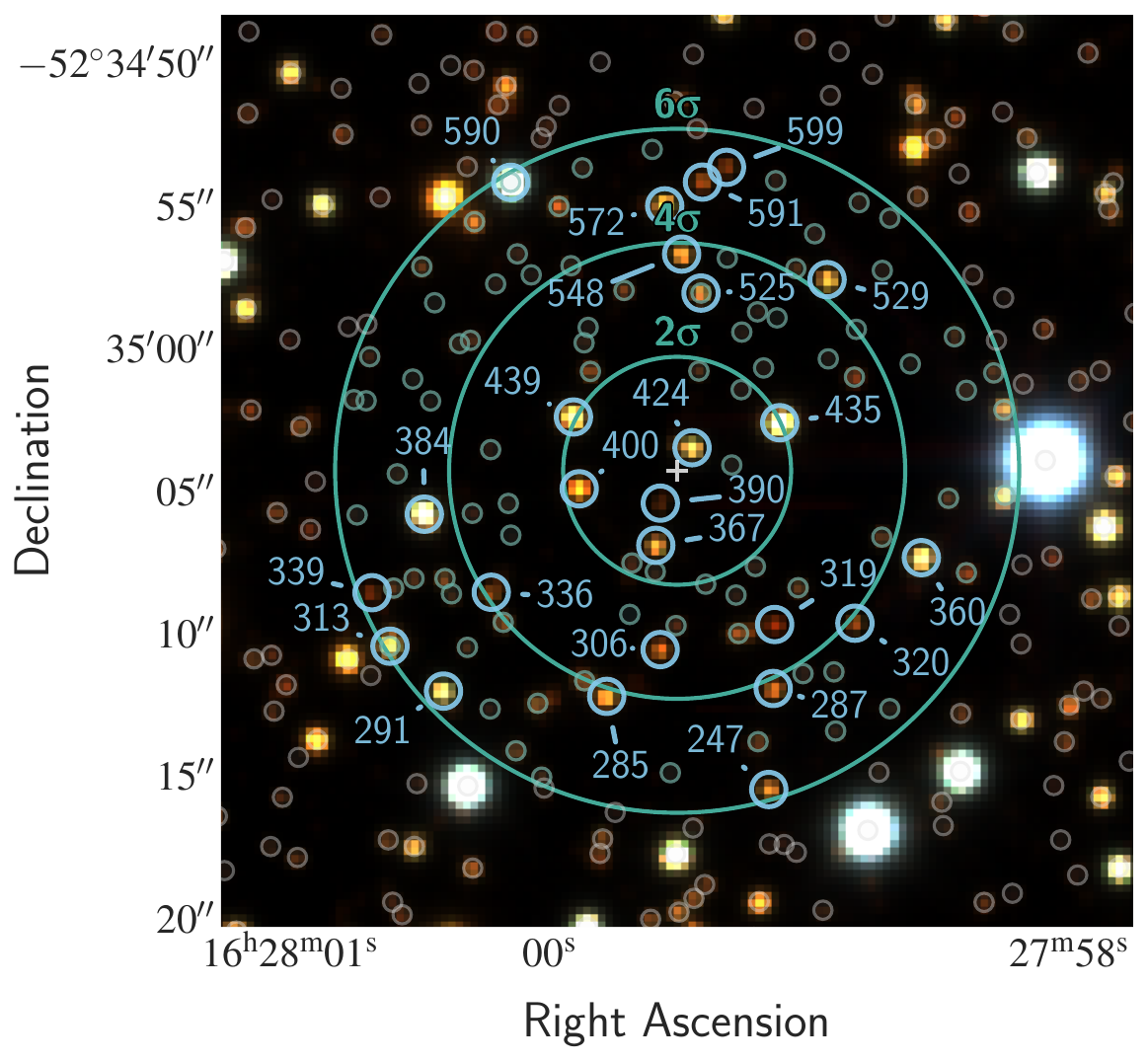}
\caption{
\label{fig:muse_fov}
The FoV and sources detected in the MUSE data in the vicinity of \gleamx\ (white cross). The pseudo-colour image was generated by collapsing the data cube in three wavelength ranges. The right panel is a zoom-in of the left (extent indicated by white box), and highlights individual sources detected by our algorithm with circle markers. The extent of the 2, 4 and 6$\sigma$ location uncertainties on \gleamx\ (where $\sigma \simeq 2$\,arcsecond) are shown with larger, labelled cyan circles, and sources within 6$\sigma$ have their circle markers coloured differently. Sources referred to throughout in text and figures, and those matched with a \textit{Gaia} source, are labelled with ids. These are arbitrary, and designated in ascending declination order from the entire MUSE cube source catalogue. The effect of 45\,degree rotations between exposures gives the FoV shown on the left as an the 8-pointed star shape.)
}
\end{figure*}

\subsection{NTT/ULTRACAM}
\label{sec:obs_nttultracam}

ULTRACAM is a high-speed, triple-beam imaging photometer \citep{dhillon2007}. A total of 682 frames of \gleamx\ were obtained with ULTRACAM on 2022 March 4, each composed of three images taken simultaneously in the Super SDSS filters $u_sg_si_s$ \citep{dhillon2021}. The instrument was used in its two-windowed, unbinned mode, with 10\,s exposure times and 0.024\,s dead time between each frame. The sky was photometric with seeing of approximately 1" and no Moon.

The ULTRACAM data were reduced using the HiPERCAM aperture photometry pipeline \citep{dhillon2021}. All frames were first debiased and then flat-fielded, the latter using the median of twilight sky frames taken with the telescope spiralling. We then shift-and-added all 682 frames to produce a deep image in each filter from which we identified 14 significant sources that fall within 12\,arcsec (6$\sigma$ localisation) of \gleamx.

\subsection{VISTA}
\label{sec:obs_vistavvv}
The footprint of the VISTA Variables in the Via Lactea (VVV) public survey \citep{vvv2010} taken on VISTA telescope serendipitously contains the localisation of \gleamx. We obtained all available VISTA/VVV $K_s$-band imaging covering the location of \gleamx\ from the ESO science portal\footnote{\url{https://archive.eso.org/scienceportal/home}}. These data are available pre-reduced \citep[for details, see][]{cross2012} and contain 99 epochs, covering a time-span from 11 July 2016 until 1 September 2019. Unfortunately, none of these data cover the period of radio activity during early 2018.

Although the baseline from early VVV epochs to our MUSE data could in principle be used to search for proper motion sources, it would require a relative astrometric accuracy of $\lesssim 0.2$\,arcsecond to probe even high Galactic transverse velocities (hundreds of \kms). Achieving this requirement is compromised by having only very few bright, isolated sources in both MUSE and VVV fields of view, alongside differences in wavelength coverage and the variable quality point spread function of the VVV data.

\section{Methods}
\label{sec:methods}

To investigate possible counterparts to \gleamx, we applied a number of methods of investigation to sources in the vicinity. For the remainder of the paper, unless otherwise stated, we will concentrate our analyses and discussion on those sources located within the 6$\sigma$ localisation uncertainty of \gleamx\ -- i.e. those sources highlighted in \cref{fig:muse_fov}. A radius encircling 6$\sigma$ of the localisation was chosen to be conservative to the presence of underestimated uncertainties, or systematics, in the radio localisation.

\subsection{Photometric variability}
\label{sec:meth_photvar}

\subsubsection{NTT/ULTRACAM}
\label{sec:meth_photvar_ultracam}

We performed photometry using normal and optimal \citep{naylor98} extraction of the 14 sources we identified, using both fixed-radius apertures and apertures that varied in radius with the seeing. The location of the sources are identified on \cref{fig:muse_fov} and detailed in Appendix~\ref{app:specific_sources}. We note two ULTRACAM sources (identified as ids 287 and 548) have comparable brightness sources in the vicinity, and were seen as blended sources. For these an aperture was used for these sources that also included flux from the neighbouring sources.

We searched for periodicities in all sources using both Lomb-Scargle \citep{press89} and
Phase-Dispersion Minimisation \citep{stellingwerf78} periodograms, the latter technique being particularly sensitive to highly non-sinusoidal signals in the optical light curves, following the methodology of \citet{dhillon2011}.

\subsubsection{VISTA/VVV}
\label{sec:meth_photvar_vvv}

We performed aperture photometry with the {\sc photutils} \citep{photutils} package on 14 detected objects within the 6$\sigma$ localisation of \gleamx -- the sources are detailed in Appendix \ref{app:specific_sources} and indicated on \cref{fig:muse_fov} (all were matched within $\sim$0.5\,arcsec of their respective MUSE source). We obtained 2MASS \citep{2mass2006} $K_s$-band magnitudes of comparison stars in each image to determine the zero point and its uncertainty. After calibrating our photometry using the zeropoint, we constructed light curves for each source, retaining those detections with a signal-to-noise ratio of at least three. Given the typical depth of a VVV image, all photometric measurements of the 14 sources was comparatively low signal-to-noise ratio. For this reason we first employed quick quality checks on the overall data to ascertain if it was conducive to more rigourous searching for variability. Each light curve was fitted with a simple constant and linear-in-time model to search for long term evolution. The residuals around each of these models were then checked to determine the presence of residuals not dictated purely by statistical noise. As will be shown in \cref{sec:res_photvar}, these results indicated no meaningful departure from constant evolution, and so we forewent further analysis of these data.

\subsection{Spectral typing and emission line search}
\label{sec:meth_spectyp}

MUSE data allows us to make a flux-limited census of the stellar sources in the vicinity of \gleamx. We use this to search for either spectral peculiarities in sources, or any systems matching theoretical expectations for the counterpart system.

In order to guide our search, we made use of {\sc PyHammer} \citep[version 2.0.0;][]{pyhammer2017,pyhammer2020} to automatically classify sources based on its own spectral template library. It additionally allowed us to visually inspect the template match for each of our source spectra -- extracted from the deep-stacked data cube -- and update the classification manually. A detailed typing was not the aim of our investigations, but rather we used the library of templates to help with broad stellar typing and anomaly detection. 
This method was performed for all sources within the $6\sigma$ localisation of \gleamx. Prior to ingestion into {\sc PyHammer}, we applied a signal-to-noise per spectral pixel cut on the spectra of 8 using the {\sc PyHammer} configuration interface.

We concentrated a search for emissions lines, and any variability in emission lines between MUSE epochs, around the \ha\ region of the spectra. \ha\ is one of the strongest lines and is readily excited, giving the best possibility to observe unusual macroscopic conditions of any counterpart system -- e.g. \ha\ is not typically excited in the atmospheres of low mass stars (although can be present during flares and heightened activity), but has been seen to be prominent in close WD binary systems, such as the WD pulsar systems, as a result of interaction or accretion processes.
For this search, the OB1, OB2 and OB3 MUSE epochs, were passed through synthetic top-hat narrow-band filter profiles: a \ha\ filter centred on 6563\,\r{A}\ with a width of 200\kms, and a continuum filter located $+12$\,\r{A}\ relative to the \ha\ filter. A width of 200\kms was chosen to comfortably capture any reasonable Doppler shifts of putative emission and the instrumental broadening of the line in MUSE data, whilst not acquiring too much added sky noise. The continuum-subtracted \ha\ filter flux was used to search firstly for significant \ha\ emission across the entire FoV, and secondly to quantify the significance of variability in the \ha\ region between the epochs. During construction of these synthetic narrow-band images, we employed the method detailed in \citet{fossati2019} and \citet{lofthouse2020} to correctly estimate variances. Succinctly, the distribution of values in sky pixels (i.e. those not flagged by our source mask used in the ZAP reduction process) divided by $\sigma$ are expected to be normal, with a scale factor of 1. $\sigma$ is calculated by propagating the pixel variances in the \texttt{STAT} extension of the MUSE data for the narrow-band image creation process. In each case, a factor 2.4--2.5 needed to be applied to the na\"ive $\sigma$ values to satisfy this condition.

\subsection{Radial Velocity variability
\label{sec:meth_rvvar}}
Many leading scenarios for \gleamx\ invoke a close binary systems containing a NS or WD. We may expect to detect such a system in our multi-epoch MUSE data as a variation in the radial velocity (RV) of a source. Based on our results of spectral typing of the sources in the vicinity of \gleamx\ (see \cref{sec:meth_spectyp}), along with the typical low SNR of our spectra, we opted to perform a full template fitting of the spectra using the {\sc spexxy}\footnote{\url{https://github.com/thusser/spexxy}} package \citep[e.g.][]{husser2016}, along with PHOENIX-ACES stellar model library \citep{husser2013} with $T_\textrm{eff} = 2300-15000$\,K, interpolated to match the resolution and the line-spread function of MUSE. The template-fitting was performed for all epochs for each source. The mid-times of our MUSE epochs were corrected to Barycentric Dynamical Time (TDB) to act as the timestamps for our RV measurements.

The MUSE instrument, in terms of wavelength-calibration (equivalently velocity accuracy) of the final data cubes, has been characterised to $\sim1$\,\kms\ over most of its wavelength coverage. However, a position-dependent shift at the level of $2.5$--$4.0$\,\kms\ for individual lines exists \citep{weilbacher2020}. Our observing strategy rotates and dithers the instrument between subsequent epochs (to alleviate detector artefacts in our stacked cubes). As such this source uncertainty manifests as another source of pseudo-random uncertainty on velocity values. We therefore add a 3.25\,\kms\ uncertainty (as the mid-point of the position-dependent uncertainty) in quadrature to our statistical uncertainty from the template fitting.
Although strategies exist to mitigate this position uncertainty, and recover the instrument accuracy \citep[e.g.][]{kamann2016}, the typical level of variability we are expecting for all but pathologically inclined systems (See \cref{sec:discussion}) would anyway dominate over this uncertainty, and so we did not include such corrections as they are not expected to impact our results.

For sources where {\sc spexxy} fitting was successful for at least 5 individual epochs, resultant RV evolution was modelled separately as a sinusoid and as a constant (i.e. a systematic offset with no evolution). The Bayesian likelihood of model parameters were explored using the nested sampling Monte Carlo algorithm MLFriends \citep{buchner2016, buchner2019}, implemented in the {\sc UltraNest} package \citep{ultranest2021} and using a Gaussian log-likelihood function. The prior on the sinusoid period was $P_\textrm{orb} \sim U(15, 300)$\,minutes to comfortably capture both the measured period of \gleamx\ at 18.18\,minutes \citep{hurleywalker2022} and the expected orbital periods of close WD binary systems. Although the expectation is that the radio emission period is not related to the binary orbit (and would be more likely the spin period of a binary constituent), we opt for this wide orbital period prior to elucidate any potentially interesting systems -- a NS harbouring system could be in a much tighter orbital configuration, for example. The semi-amplitude and systematic offset priors were $K_2 \sim U(0, 900)$\,\kms and $\gamma \sim U(-600, 600)$\kms, respectively to, again, comfortably encompass any potential system. The same systematic offset prior was used for the constant model. 

Although baseline and coverage of our MUSE data are not ideal for a robust search for $P_\textrm{orb}$ on timescales of hours, such systems should have $K_2$ values of hundreds of \kms\ \citep[e.g.][]{marsh2016}, meaning a modelled search combined with visual inspection would reveal these readily in our data. There is, however, always the potential for binary-plane inclination to diminish the expected amplitude.

We finally note that individual MUSE exposure data cubes are susceptible to significant artefacts, brought on by the image slicing instrumentation configuration. Although typically mitigated by stacking of exposures, here it adds an additional source of variable noise within the individual epochs. This accounts for some individually discrepant points, and the variable precision of the RV measurements within a pointing. This also did not allow simple statistical cuts (e.g. on $\chi^2$/dof) to be trusted at face value, and we instead relied on manual inspection alongside our sinusoid fitting procedure.

\section{Results}
\label{sec:results}

In \cref{fig:muse_fov} we indicate ids for specific sources that were further investigated or are referred to individually in the following section.

\subsection{Photometric variability}
\label{sec:res_photvar}

\begin{figure}
\includegraphics[width=\linewidth]{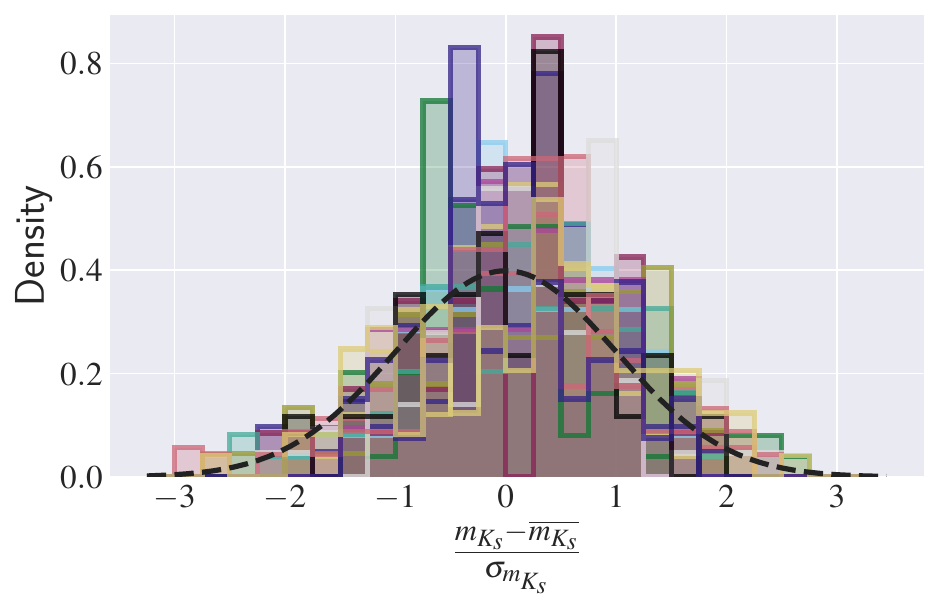}
\caption{
\label{fig:vvv_residuals}
Sigma-normalised residuals of individual $K_s$-band photometry detections from a constant model ($\bar{m_{K_s}}$) in the light curves of the five VISTA/VVV sources investigated. The lack of significant deviation beyond that expected from statistical noise (i.e. a normal distribution centred on zero with a standard deviation of one, indicated by the black dashed line) from the constant model indicates consistency with the model and the inadequacy of the data to perform more detailed searches of periodic behaviour.
}
\end{figure}

In \cref{fig:vvv_residuals} we plot the residuals of individual photometric detections in our VISTA/VVV source light-curves from the constant model. All data points for each source are consistent with being distributed around a constant model within the uncertainties of the photometry, therefore precluding a deeper investigation into the nature of any variability of the sources. Fits using a linear-in-time model for the light curves gave slopes broadly consistent with no evolution, given the accuracy of the data. Much deeper NIR data will be required to place meaningful constraints on the nature of any variability in this regime.

We additionally found no evidence for periodic or other variability (via manual inspection) in any of the 14 ULTRACAM sources within (6$\sigma$) localisation of \gleamx.

\subsection{Spectral typing and emission line search}
\label{sec:res_spectyp}

The majority of spectra precluded even broad stellar typing owing to relatively low signal-to-noise ratio. For $\sim$70\% of sources, neither the spectral match provided initially by {\sc PyHammer}, nor manual inspection, provided a reasonable match. These sources were typically red continua with an absence of features (at the level of our data quality), although there are a small number of brighter sources with more clearly featureless red continuum spectra. For those sources where typing was possible, the population consists of predominantly early M-dwarfs and late K- or G-type stars. Three spectra of bright stars in the vicinity were presented by \citet{rea2022}, these are source ids 424, 400 and 439, labelled by those authors G1, G3 and G4, respectively. We confirm comparable spectral typing and the absence of distinguishing features, as noted by \citet{rea2022}, for these sources. Importantly for the model of \citet{loebmaoz2022}, we find no indications of a hot sub-dwarf local to \gleamx -- i.e. no source spectra displays a blue continuum with strong H absorption lines. Any such hot sub-dwarf at a distance of $\sim1.3$\,kpc would be well-detected in our data (see \cref{sec:discussion}).

Even the brightest WDs would have an apparent visual magnitude of $\gtrsim21$\,mag at the distance of \gleamx, with more typical WDs being $\sim22-24$\,mag \citep{gentilefusillo2021}. As such the prospects of directly identifying any WD counterpart spectroscopically in our MUSE data is limited. This issue is compounded by the fact that any viable WD counterpart candidate system would be in a tight binary system \citep[][although see \citealt{rea2024}]{katz2022} and so is likely to be outshone significantly at our wavelength coverage ($\gtrsim4800$\,\r{A}) by its non-degenerate companion. This is also demonstrated in \citet{hurleywalker2024}, where optical spectra of the M3V optical counterpart for a comparable distance long-period radio transient (GLEAM-X\,J0704-37) precludes detection of a companion WD with similar wavelength coverage.
The lack of a spectral-typed WD from the data did not therefore strongly constrain the presence of a WD in the localisation region of \gleamx, which is better probed by variability information (\cref{sec:res_photvar,sec:res_rvvar}), as well as the presence of peculiar spectral features in a companion. 

\begin{figure}
\includegraphics[width=\linewidth]{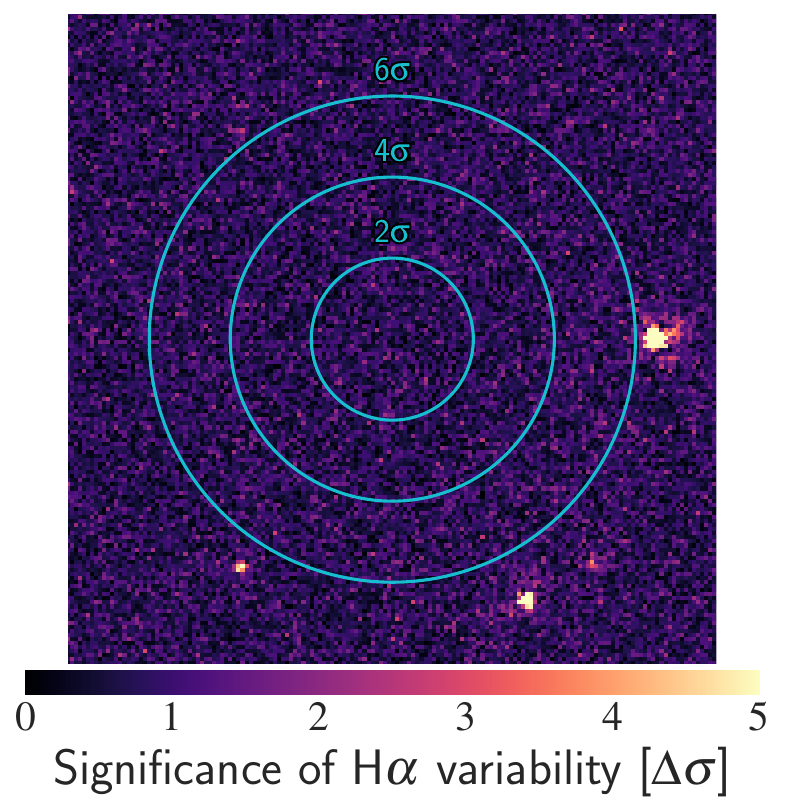}
\caption{
\label{fig:muse_havar}
The maximal variation of \ha\ emission between the three OB-stacked MUSE cubes, expressed as a sigma value -- i.e. the significance of \ha\ variability -- with localisation contours of \gleamx\ overlaid. No significantly varying source is found within 6$\sigma$ of \gleamx. Sources around the edge of the frame are residuals from very bright stars owing to imperfect flux calibration.
}
\end{figure}

Using continuum-subtracted \ha\ narrow-band images we found no significant variability of any source within the 6$\sigma$ localisation of \gleamx\ -- the most significant individual pixel in this region shows a variability of $\sim$3$\sigma$ in flux, with no signs of an astrophysical source undergoing variation. \cref{fig:muse_havar} shows the significance of spaxel variability in the \ha\ spectral regime. For the three prominent sources outside the localisation of \gleamx\ we see no evidence of particularly unusual behaviour, and indeed, given they are the three brightest sources in the field, ascribe the variation seen as due to flux-calibration uncertainties. 

\begin{figure*}
\includegraphics[width=\linewidth]{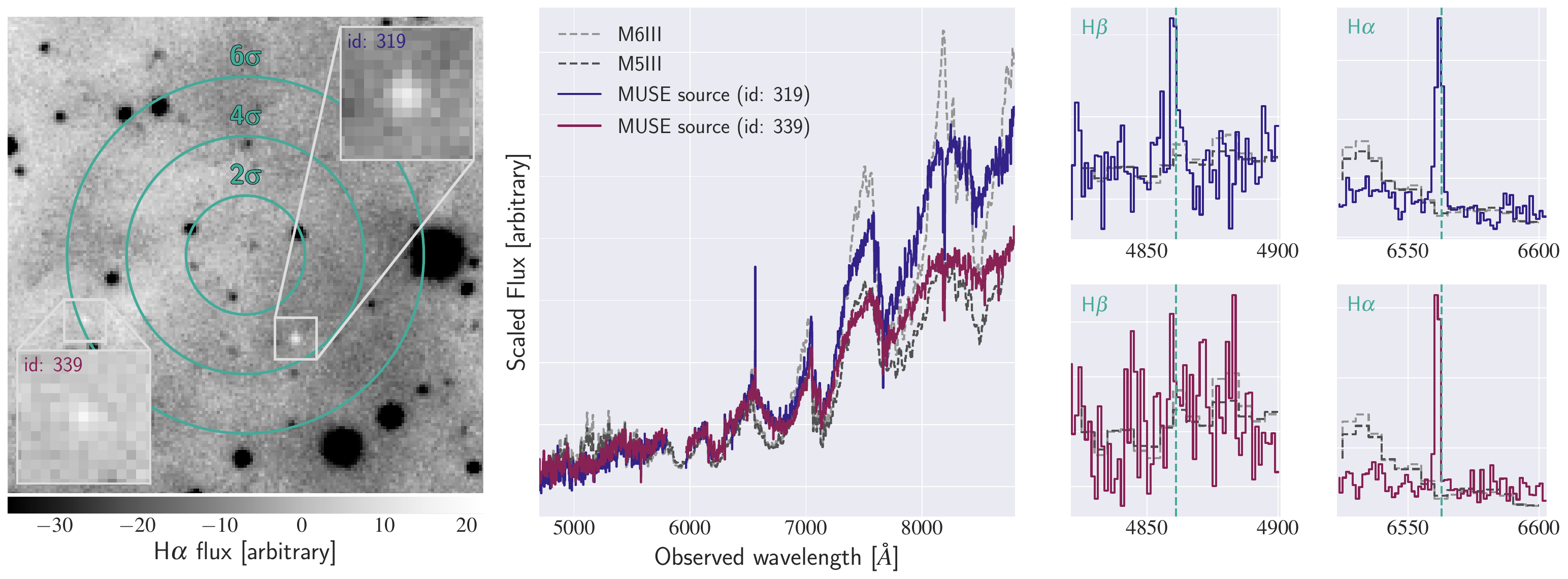}
\caption{
\label{fig:muse_hasrc}
\textit{Left:} The mean \ha\ narrow-band emission image from OB1, OB2 and OB3 MUSE epochs. Two sources are seen in emission, approximately 3.2 and 6.0$\sigma$ from the localisation of \gleamx. \textit{Centre:} The spectra from the `all' MUSE epoch of the emission sources. Shown in comparison are spectra of M5III and M6III stars \citep{pickles1998}. Spectra were normalised to the same median flux level in the $4800-6800$\,\r{A} wavelength range. \textit{Right:} A close-up of the \ha\ and \hb\ regions of the spectra, with the MUSE source spectra shown on different subplots for clarity. The RV of these sources revealed no evidence for a short-period binary configuration (see \cref{sec:res_rvvar}).
}
\end{figure*}

Following a search for variable \ha\ emission, a visual inspection of a mean continuum-subtracted \ha\ image (formed from those of OB1, OB2 and OB3) highlighted two point-sources of emission in the vicinity of \gleamx: source ids 319 and 339. The source spectra display clear unresolved \ha\ emission and, in the case of source id 319, \hb\ emission (\cref{fig:muse_hasrc}). A comparison of the stellar spectrum with stellar templates from the ESO MUSE Stellar Library \citep{ivanov2019} and the Pickles Stellar Library \citep{pickles1998} was performed, indicating they are approximately M5-M6 type stars, which was also corroborated by our initial {\sc PyHammer} typing. Radio emitting close WD binary systems such as AR Sco \citep{marsh2016} and J191213.72-441045.1 \citep{pelisoli2023}, display prolific \ha\ in emission on top of a M-dwarf spectrum, due to irradiation of the WD companion in each system. However, other non-binary mechanisms for \ha\ emission in M-dwarfs are also possible, as discussed later.

The equivalent width (EW) of \ha\ emission for source ids 319 and 339 are given in \cref{tab:haew} for each of our combined epochs. There is no evidence for variation in the \ha\ strength for a given source.

\begin{table}
    \centering
    \begin{tabular}{ccccc}
        \hline
        {id} & {OB1} & {OB2} & {OB3} & {Mean} \\
        \hline
        319 & $-5.2 \pm 0.7$ & $-7.4 \pm 0.8$ & $-5.9 \pm 0.8$ & $-6.1 \pm 0.4$ \\
        339 & $-4.8 \pm 0.8$ & $-3.1 \pm 0.9$ & $-3.6 \pm 1.0$ & $-3.9 \pm 0.5$ \\
        \hline
    \end{tabular}
    \caption{\ha\ equivalent widths for the two emission lines sources in the vicinity of \gleamx\ along with their weighted mean over the three epochs. Values are in \r{A}.}
    \label{tab:haew}
\end{table}

\subsection{Radial Velocity variability
\label{sec:res_rvvar}}

\begin{figure*}
    \centering
    \includegraphics[width=\linewidth, clip, trim=0 48.5 0 0]{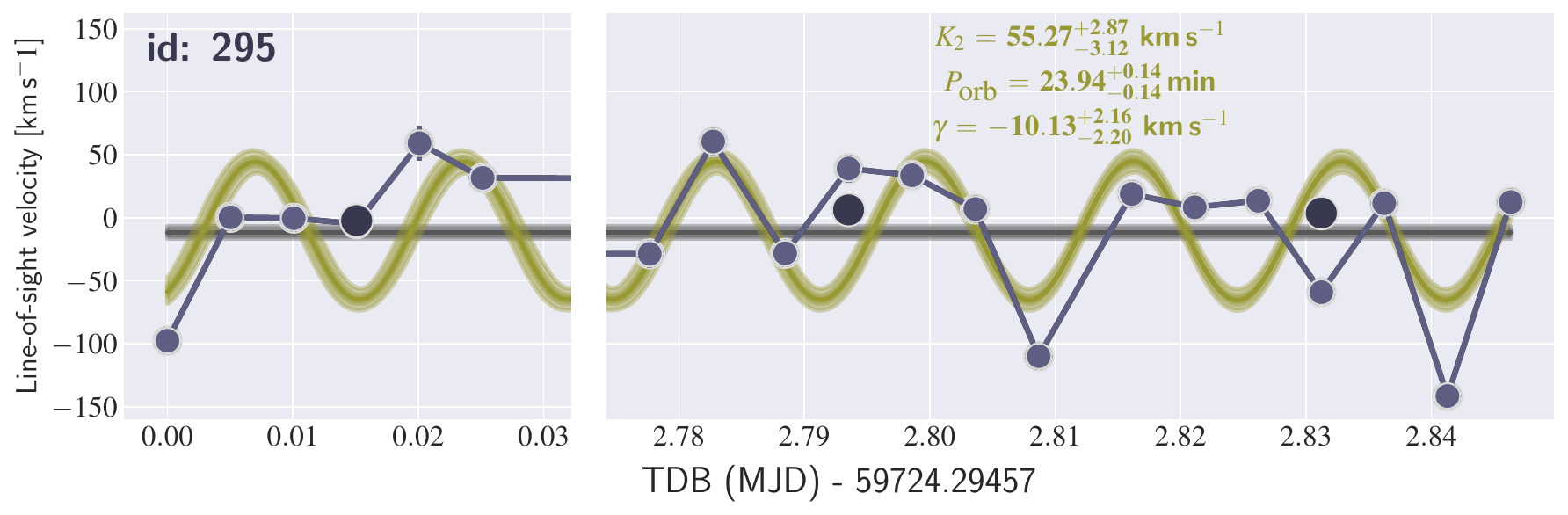}
    \includegraphics[width=\linewidth, clip, trim=0 48.5 0 0]{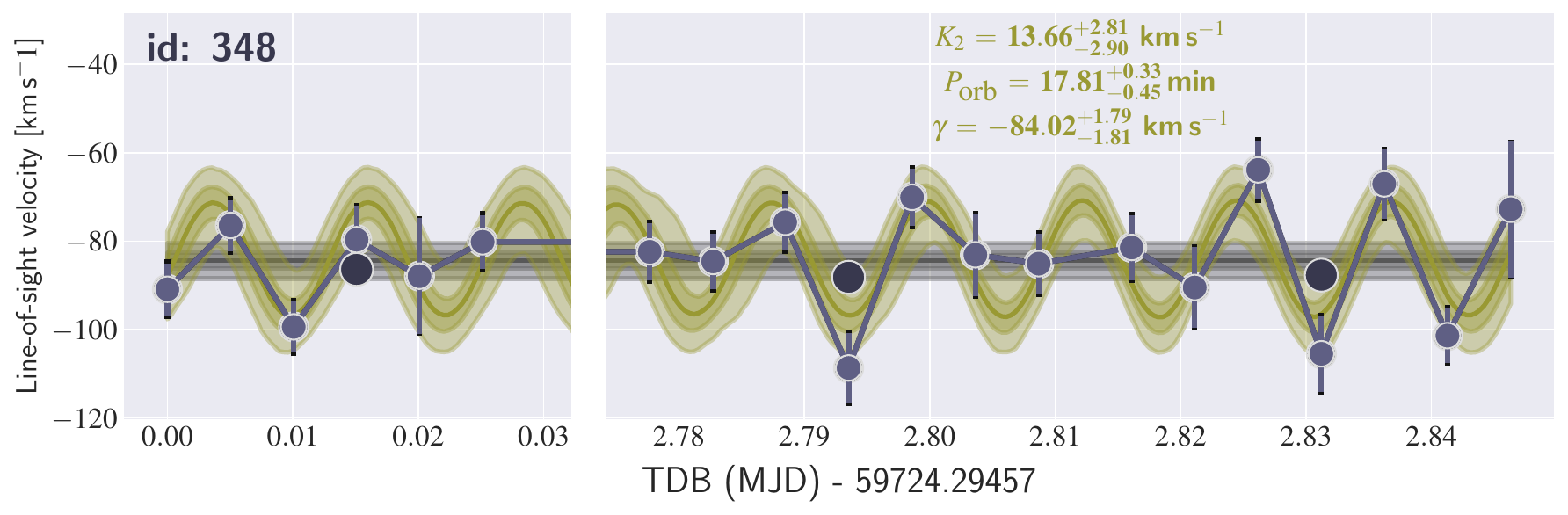}
    \includegraphics[width=\linewidth]{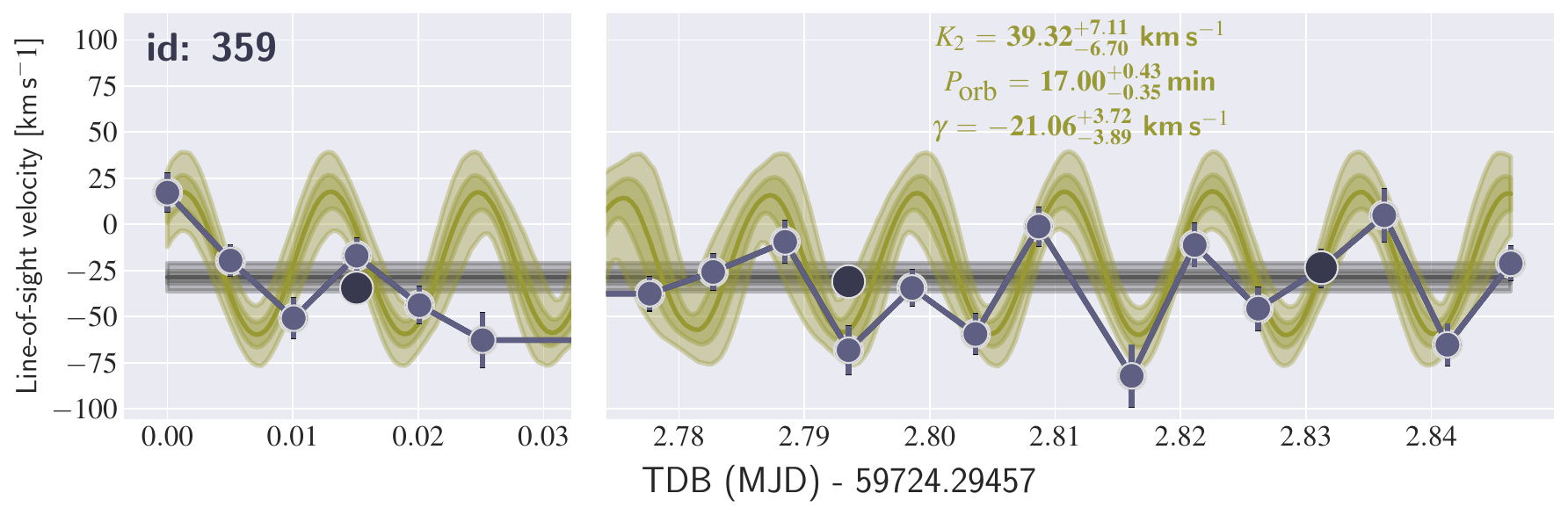}
    \caption{Radial velocity measurements for sources passing our initial vetting cuts for sinusoidal fitting. We show these sources as the examples with a closest approximation to a tight binary signature expected from many models of \gleamx. However, we make no strong claims on their reality as such systems owing to: the poor description of the data by the sinusoid in some cases, the possibility of underestimated/unaccounted for uncertainties on individual measurements, and the astrophysical implications of such detections (see text). Measurements from individual MUSE epochs are shown as lighter circle markers, joined by solid lines, with larger, darker markers indicating the measurements from the combined epochs OB1, OB2, and OB3. Bars on each marker show the statistical uncertainty from {\sc spexxy} alone, and for the increased error budget including instrumental effects in black (see \cref{sec:meth_rvvar}), although not always visible. On each subplot, the median posterior models are shown by thick lines in green and grey for the sinusoid and constant models, respectively. The $P_{16-84}$ and $P_{1-99}$ percentile regions for each model are shown by the shaded regions. Sinusoid parameters are given as the median and $P_{16-84}$ percentile interval. Locations of sources (based on their id) are indicated on \cref{fig:muse_fov}.}
    \label{fig:rvvar}
\end{figure*}

\begin{figure*}
    \centering
    \includegraphics[width=\linewidth, clip, trim=0 48.5 0 0]{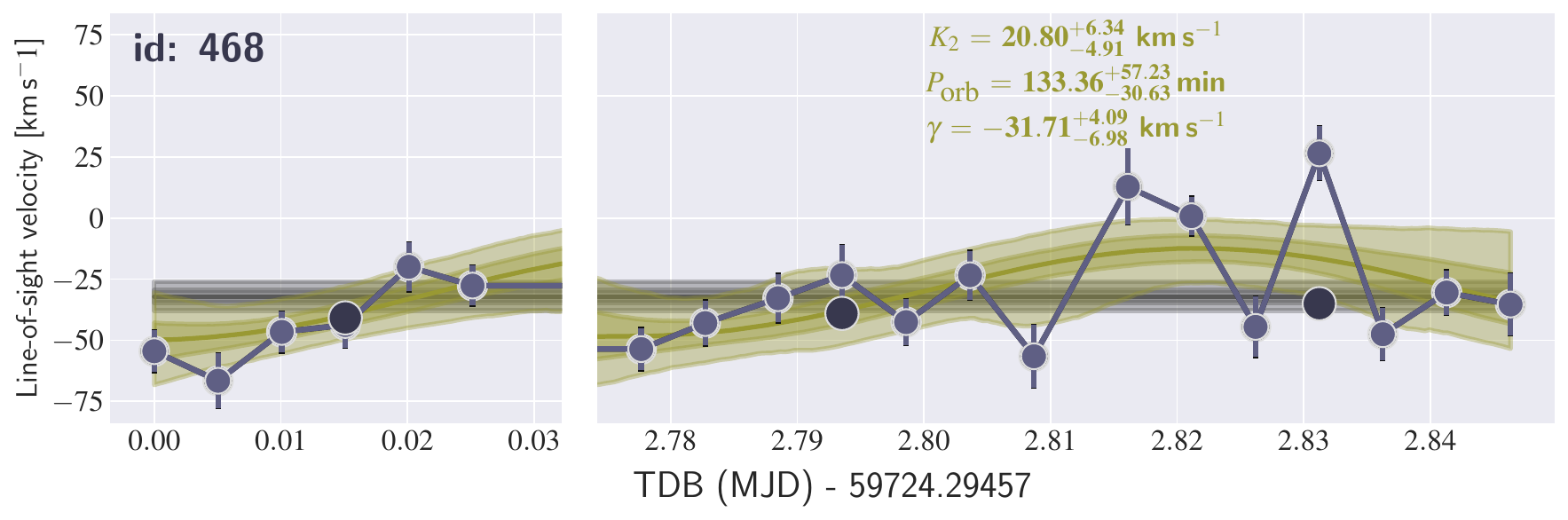}
    \includegraphics[width=\linewidth, clip, trim=0 48.5 0 0]{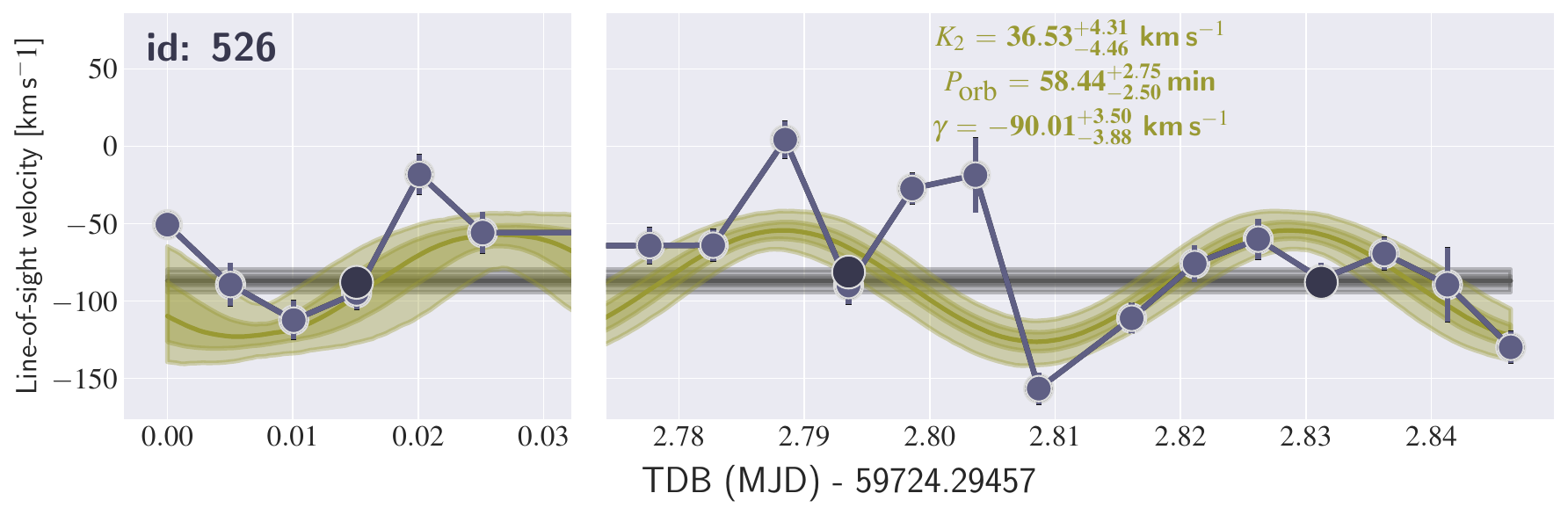}
    \includegraphics[width=\linewidth]{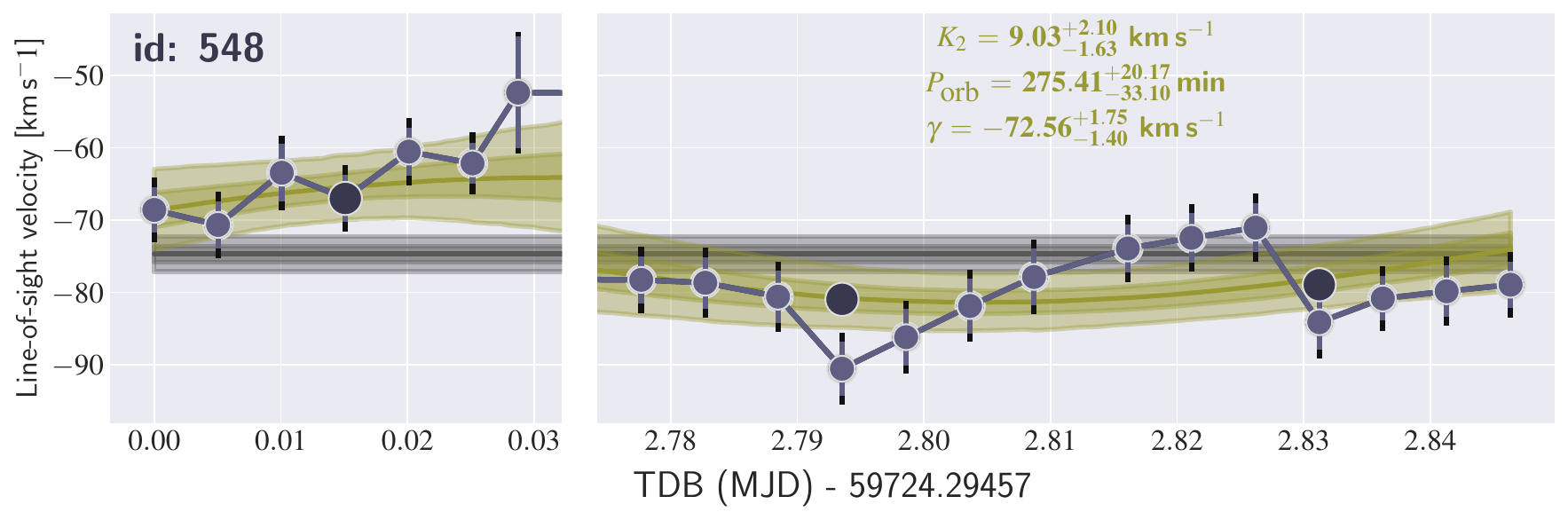}
    \caption{Same as \cref{fig:rvvar}, but here shown for three other sources with model fits alluding to longer RV periodicity.}
    \label{fig:rvvar2}
\end{figure*}

Following the method detailed in \cref{sec:meth_rvvar}, there were 78 sources in the 6$\sigma$ region of \gleamx\ for which we could compute sinusoid and constant model parameter posteriors. These models were visually inspected for each source. Sources were discarded that had either a) a Bayes Factor indicating no strong preference for the sinusoid model over a constant evolution, b) no well-determined period and amplitude, c) discrepant individual measurements driving the posteriors or d) RV measurements from the OB1, OB2 and OB3 epoch spectra that were not representative of the mean of the individual epoch RVs for each -- this latter check was effective to discriminate against low SNR spectra for which variations in individual epochs were driven by the limitations of the template fitting technique. Sources that pass these inspection criteria are shown in \cref{fig:rvvar,fig:rvvar2}. 

Each source displays some form of variability that appears inconsistent with no evolution (i.e. the constant model), although in cases the sinusoid fits also do not present a convincing description of the data. Three sources (ids 295, 348 and 359; \cref{fig:rvvar}) have solutions favouring $P_\textrm{orb} \lesssim 0.4$\,hr. If these were to be real, they would be tight binaries in the regime of compact binaries, and would additionally require very low orbital inclination to explain the comparatively low semi-amplitude of the sinusoid (see \cref{sec:discussion}).
For three other sources (ids 468, 526 and 548; \cref{fig:rvvar2}) we find $P_\textrm{orb} \sim 2,1,4.5$\,hr, respectively. Sources 526 and 548 are comparatively bright, and yet their RV curves appear erratic with respect to the sinusoid model. Similarly to sources fitted with shorter periods, the semi-amplitudes seen are low compared to that expected for realistic binary configurations in all but low orbital inclination systems. As elaborated further in \cref{sec:discussion}, although the variations in RV do not appear to be simply systematic variations between epochs for the different sources, we do not consider the above six systems as strong binary candidates due to astrophysical considerations, not least because of the huge inferred spatial density of such systems.
In terms of localisation, source ids 359 and 468 are within the 2$\sigma$ localisation of \gleamx.
Where determined, the RV curves of the other sources within 2$\sigma$ of \gleamx\ are additionally shown in Appendix~\ref{app:2sig}.

\begin{figure*}
    \centering
    \includegraphics[width=\linewidth, clip, trim=0 48.5 0 0]{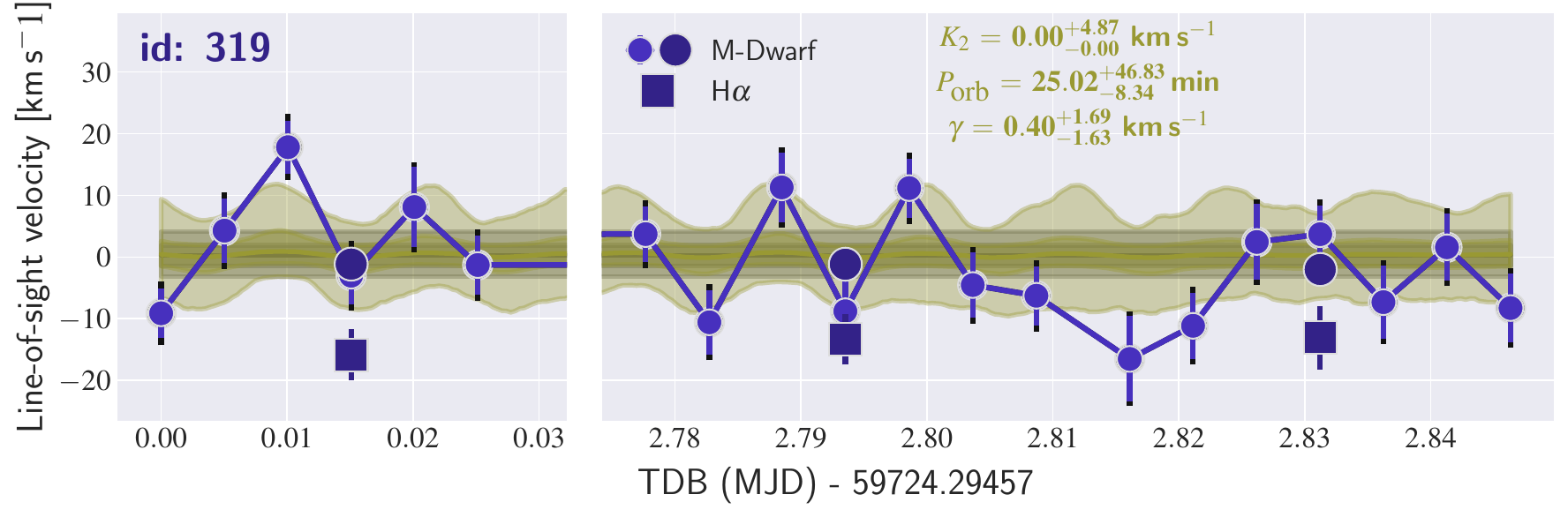}
    \includegraphics[width=\linewidth]{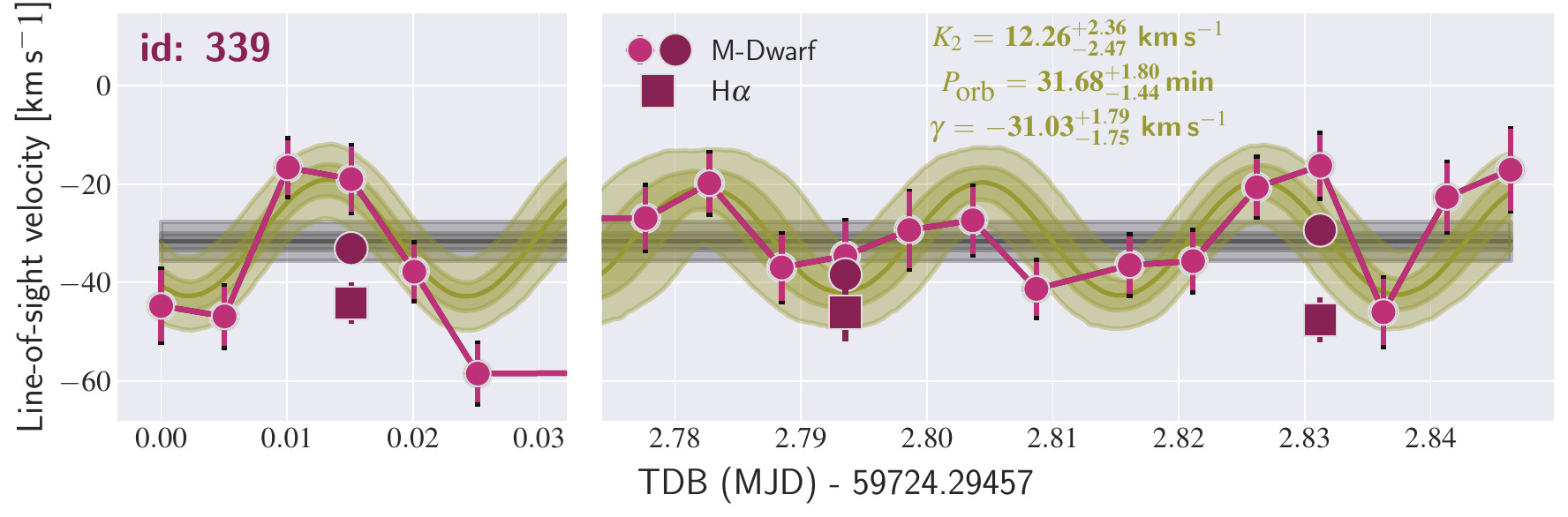}
    \caption{Same as \cref{fig:rvvar}, but here for two emission line sources identified in \cref{sec:res_spectyp} and shown in \cref{fig:muse_hasrc}. Neither show strong evidence for RV variation and such low $P_\textrm{orb}$ values as favoured by our posterior sampling are unrealistic for a binary containing a non-degenerate star. Also shown is the RV of the \ha\ emission lines are show for the combined OB1, OB2, and OB3 epochs as large square markers. Sources are colour-coded the same as \cref{fig:muse_hasrc}.}
    \label{fig:emissionrvvar}
\end{figure*}

We further analysed the two emission line sources identified in \cref{sec:res_spectyp} by comparing the radial velocities from the {\sc spexxy} template-fitting package to characterise the M-Dwarf stellar spectrum, with that of an emission line fit to \ha. The fit to \ha\ was done using a single Gaussian and a 3rd order Legendre polynomial for the local continuum. Sampling of the posterior distributions used a similar {\sc UltraNest} procedure as used for the RV modelling (\cref{sec:meth_rvvar}). The results are shown in \cref{fig:emissionrvvar}. For source id 319 no RV variation is found, whereas formally the sampling determines a low-amplitude, short-period for source id 339 but with no well-defined phase. As above, for reasons discussed in \cref{sec:discussion}, we do not consider this RV evolution to be indicative of expected counterpart behaviour. Noticeably, there appears to be an offset between the line-of-sight velocity determined from the template-fitting and the \ha\ emission line fit. Line emission in binary systems may arise from geometrically distinct locations compared to the centre of light and so be offset in velocity. In isolated M-dwarfs, it would require a localised region of emission in order to show such an offset, but in this case we would also expect its velocity offset to change with time due to stellar rotation. As the level and direction of offset is similar and unchanging between the two sources, we conclude it is unlikely to be astrophysical in origin. Although the underlying reason for this remained uncertain after investigation,\footnote{A comparison of the velocities of \ha\ absorption, determined equivalently, and {\sc spexxy} template fits for the bright source id 435 did not show this systematic offset.} the constant offset with epoch in each suggests the emission is arising from the star in each case.

\section{Discussion}
\label{sec:discussion}

The location of \gleamx\ is in a dense region of the Galactic plane. Our stellar typing using MUSE spectra of sources in the localisation showed the population to be composed primarily of low-mass main-sequence stars, as is to be expected. Within this population we found no evidence of isolated WDs or hot subdwarf stars. In order to quantify our sensitivity to such sources we performed a photometric analysis of our MUSE data. The wavelength coverage of MUSE is comparable to the \textit{Gaia} \textit{G}-band, excepting a lack of coverage from $4000-4700$\,\r{A}. We built a white-light image from our ``all'' epoch and cross-matched aperture photometry of sources detected using {\sc sep} \citep[][itself built on the algorithm of \citealt{sextractor}]{sep} with the \textit{Gaia} Data Release 3 \citep{gaiadr3} catalogue, accessed via {\sc astroquery} \citep{astroquery}. From this we estimate a zero point of 28.1\,mag and a 5$\sigma$ limiting magnitude of 24.4\,mag in the white light image. Spectral typing was possible $\sim1$\,mag brighter than this limit at the level of identifying the spectral continuum shape. As discussed in \cref{sec:res_spectyp}, WDs are expected to be $\sim22-24$\,mag at the distance of \gleamx. Therefore, although we would expect to have been able to highlight WD candidates based on a blue-continuum slope, we would not expect to make a robust classification. Nevertheless, as shown in Appendix~\ref{app:2sig}, no detected source within 2$\sigma$ of the localisation shows any characteristics of a WD considering the expected modest extinction along the line of sight to \gleamx. Should the distance to \gleamx\ be revised significantly downwards, as was the case for GLEAM-X\,J0704--37 \citep{rodriguez2025}, our limits become more constraining on the presence of a WD, although any significant revision in distance is likely to require a multi-wavelength counterpart in the first place. Unlike WDs, however, hot sub-dwarf stars are comparatively bright in optical -- $M_G\sim 2-7$\,mag \citep{culpan2022}. This would mean the faintest would appear at $m_G\sim17.5$\,mag even at the upper distance estimate for \gleamx. As such we rule out the presence of a counterpart matching the expectations of \citet{loebmaoz2022}. The presented data also constrains short- and long-timescale brightness changes (ULTRACAM and VVV, respectively) for the brighter sources in the localisation, whereas variability has been seen in counterparts to other long-period radio sources \citep[e.g.][]{pelisoli2023}.

Close binary systems including a WD provide us with additional plausible counterpart systems to the radio emission given such systems display indicators of the close interaction between a WD and a non-degenerate companion in the form of emission lines. Although we identified no variable \ha\ emission, two point sources of constant \ha\ emission were identified (source ids 319 and 339, \cref{fig:muse_hasrc}). The configuration of their spectra is reminiscent of the known WD pulsars \citep{marsh2016, pelisoli2023} -- namely Balmer emission superposed on a mid M-Dwarf spectrum. In the case of WD pulsars, the Balmer emission is seen to be both strong and variable -- \ha\ EW varying from tens to $>100$\,\r{A} across their orbital phase \citep{garnavich2019, pelisoli2023} -- due to WD-irradiation of a face of the M-dwarf. Indeed, even for detached (non-interacting) close M-dwarf--WD systems, such irradiation produces strong Balmer emission \citep{marsh1996}, unless the WD is very cool.
Our results show no strong evidence of variability in the \ha\ EW of our sources (\cref{tab:haew}), and in any case they are at a much lower level ($<10$\,\r{A}). Also unlike the WD pulsars' spectra, we did not detect helium emission in our sources. (A search at the strongest optical transition, \ion{He}{1} $\lambda5876$\,\r{A}, was compromised due to the notch cut from our spectra by the {Na\,D} guide laser on MUSE, however.)
Balmer emission in M-dwarfs can however be produced by isolated stars due chromospheric activity. The characteristics of this emission, and particularly its relation to the rotation period of the star, at least up to the point of saturation, points to a magnetic dynamo origin \citep[e.g.][and references therein]{reiners2012,newton2017,bustos2023}. Our \ha\ EW measurements are well within the expectations of M-dwarf activity \citep{newton2017,kumar2023}.
These sources also display no strong RV periodicity indicative of a close binary system (\cref{fig:emissionrvvar}), which should have a significant amplitude except for face-on systems.
Finally, our spectra of the sources do not show any blue-continuum contribution above that expected from the M-dwarf template (\cref{fig:muse_hasrc}). Such a slope would be indicative of the presence of a WD, although we reiterate our MUSE data are only sensitive to the relatively luminous WDs at the distance of \gleamx\ and in any case do not extend to sufficiently blue wavelengths where a WD would contribute most flux to the system.
For the above reasons, we do not ascribe either of these emission sources as being credible counterparts on the available information.

Plausible systems for \gleamx\ and other long-period radio transients are close binary systems. Although how (or whether) the period of radio emission relates to the orbital period of each system can be less clear. The recently discovered ILT J$1101+5521$ system \citep{deruiter2024} has shown radio emission can occur on the orbital period of a system (likely due to magnetic locking of the rotation of the WD), but the radio period in that case is $\sim2$\,hours. In other systems with radio periodicities of minutes and identified counterparts, the period is linked to the spin of the degenerate companion \citep{marsh2016, pelisoli2023}, while the systems are in a binary orbit with $P_\textrm{orb}\sim$\,hours. The 18.18\,min radio period seen for \gleamx\ is not expected to be directly related to a binary orbit involving a WD or NS and a non-degenerate companion (the companions our data are sensitive to), as such a system would not be stable. Succinctly, any plausible system with an orbital period at the radio period of \gleamx\ would require a compact binary system (WD/NS + WD/NS), which are beyond the sensitivity of our data for RV analysis.
We preface the next discussions on our RV analysis by emphasising we do not find any convincing signature of a short-period binary system and so no strong counterpart candidate, but provide a full discussion for completeness. 

\begin{figure}
    \centering
    \includegraphics[width=\linewidth]{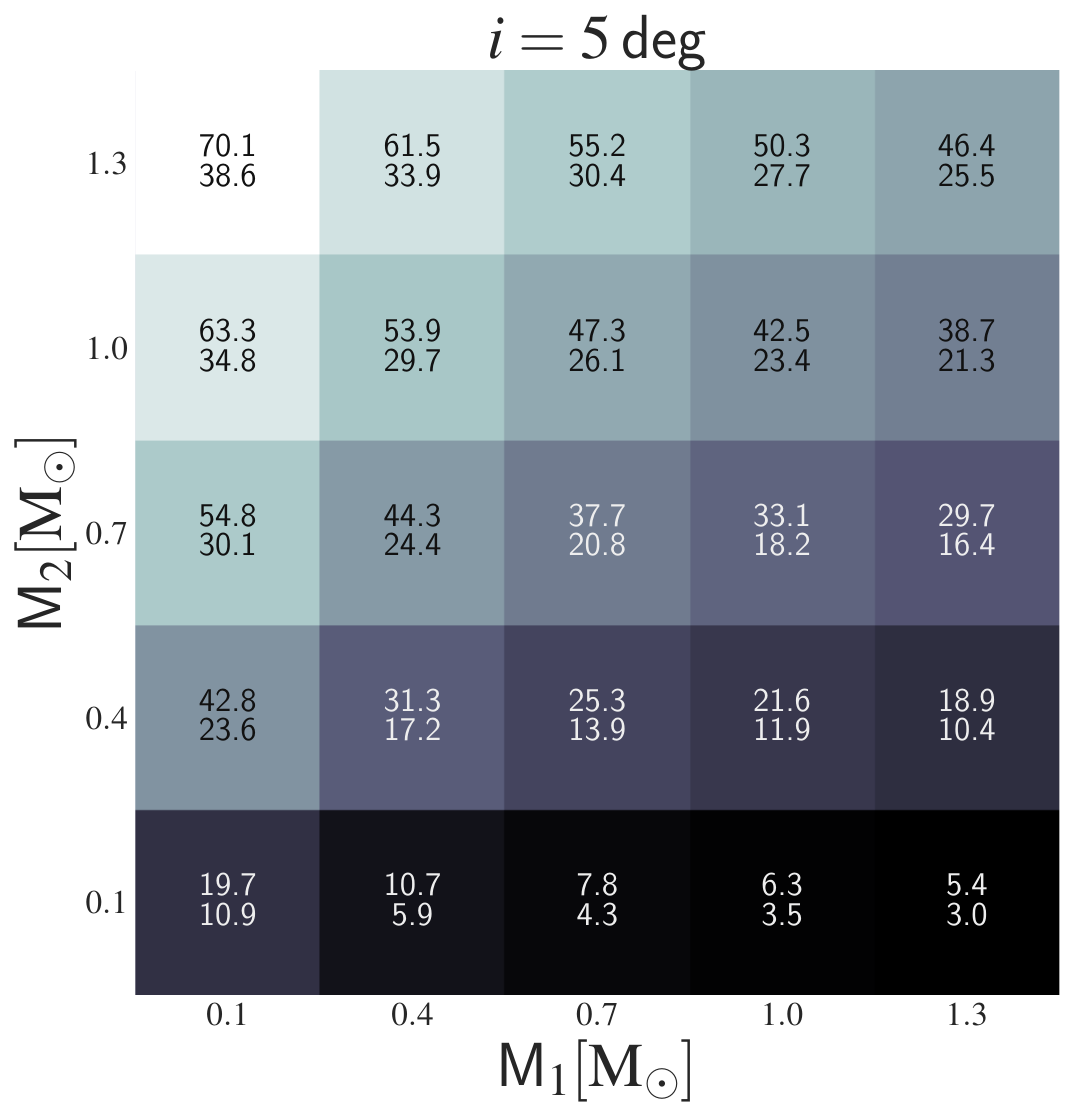}
    \caption{RV semi-amplitude ($K_2$) values for a series of $M_1$ and $M_2$ configurations according to the binary mass function. The values are in \kms\ and are given for $P_\textrm{orb} = 0.5, 3$\,hr for top and bottom values in each cell, for representative binary orbits.}
    \label{fig:bmf_rv}
\end{figure}

Our search for sinusoidal RV behaviour, indicative of a binary system, confirmed no source provides a convincing candidate with $P_\textrm{orb}\lesssim0.5$\,hour, in line with astrophysical arguments above. What variability was seen has semi-amplitudes of $\lesssim55$\,\kms. We can constrain the expected amplitudes of RV variability with typical expected binary configurations using the binary mass function:

\begin{equation}
    \frac{(M_2 \sin{i})^3}{(M_1 + M_2)^2} = \frac{P_\textrm{orb} K_2^3}{2\pi G},
\end{equation}

where $M_1$ and $M_2$ are the seen and unseen component masses respectively, $i$ is the orbital inclination, $P_\textrm{orb}$ is the orbital period, $K_2$ is the radial velocity semi-amplitude and $G$ the gravitational constant. This is shown in \cref{fig:bmf_rv} for a modest inclination of $i=5$\,deg. At a very modest orbital inclination of 5\,deg, any unseen component mass typical of a NS ($\gtrsim1.0$\,M$_\odot$) would produce a RV semi-amplitude of $\sim50$\,\kms at $P_\textrm{orb} = 0.5$\,hour. Although comparable to the amplitude seen for source id 296 (\cref{fig:rvvar}), as mentioned, such a short period is in any case not expected to be stable for a non-degenerate companion. Additionally the RV variation is not convincingly sinusoidal, and the source being 4$\sigma$ from \gleamx. Although source id 359 is within 2$\sigma$ of the localisation, it would require a low mass NS in a pathologically inclined orbit, even notwithstanding that the inferred $P_\textrm{orb}$ is unrealistic for a stable system. Our search for evidence of longer period sources with $P_\textrm{orb} \sim$\,hours, as shown in \cref{fig:rvvar2}, again revealed no convincing candidate. What RV amplitudes are favoured by the posterior sampling is at very low amplitudes, similarly requiring very low orbital inclination systems. We further note other sources of RV variability could be contributing that are unaccounted for, in particular stellar activity can reach imprints of several \kms for the typical mid-type M-dwarf, reach above 10\kms for late-types \citep{jenkins2009}.

Low mass stellar companions with a typical $\sim0.6 M_\odot$ WD companion offer some more flexibility with respect to plausible inclinations. On the balance of proximity and data quality, source id 468 perhaps best approximates to the signature one would expect, but even this source remains unconvincing based on available data.
For  $P_\textrm{orb}\simeq$\,hours WD-hosting binaries with $i=5$\,deg, one would still expect a semi-amplitude of $>10$\,\kms (\cref{fig:bmf_rv}), excepting extremely low mass WD companions \citep[which, although capable of housing significant magnetic field strengths as per][would require relatively rare products of binary evolution]{hardy2023}. Longer baseline and additional measurements are be needed to properly rule on the presence of RV variability in sources we have shown here, with the expected signatures being in the grasp of such observations.
Only one source analysed as part of our RV investigations was detected in our searches for photometric variability in the vicinity of \gleamx, source id 548, whose ULTRACAM and VVV light curves are shown in \cref{app:lightcurve}. ULTRACAM data were found to harbour no obvious periodic features following our analysis in \cref{sec:meth_photvar}. The source is at the limit of VVV data (typically a 5$\sigma$ detection per epoch) with a $K_s = 18.4$\,mag. The light curve displays no variability beyond statistical noise, in line with our results from all VVV sources in \cref{sec:res_photvar}.
A final note is to state that on spatial density alone, one would not expect such a number of short-period binary systems in a search of 0.125\,arcmin$^2$, further casting doubt on their origin.
Overall, we do not consider our measured RV curves to show any strong evidence of a counterpart system.

When considering a non-binary source system, \citet{rea2024} suggest that isolated magnetic WDs emitting beamed dipole radiation are a more attractive proposition to explain sources such as \gleamx\ than neutron stars, at least in terms of expected numbers and so constraints on their presence are of interest (in the binary scenario, they will be outshone by their companion). As those authors additionally note however, the known population of isolated, magnetic WDs fall significantly short of the requirements for radio emission in the pulsar scenario (typically by a factor $\gtrsim$100 in magnetic field strength), and so some form of binary interaction remains favoured for their interpretation. For identifying nearby isolated WDs, our observations largely do not reach the required sensitivity.

An isolated (or otherwise wide-binary) magnetar interpretation remains viable with our lack of a counterpart detection. Indeed, spectra of such objects may appear as red, largely featureless, spectra \citep{hare2024}, which are abundant in the vicinity of \gleamx\ (\cref{fig:spec2sig}). Although the optical and NIR emission of magnetars is relatively poorly understood, largely due to the difficulty in observations \citep{mignani2009, chrimes2022}, they are known to be subtly variable on short timescales linked to their spin period \citep{dhillon2011}, as well as more dramatically on longer timescales of years \citep{lyman2022}. The origin of this longer timescale emission is unclear, but, nevertheless, monitoring of this location with deep NIR observations over a significant baseline could reveal any such long-timescale variability to identify a putative magnetar counterpart.

A multi-wavelength counterpart of \gleamx\ remains elusive, despite extensive efforts to find it, including this work. The comparatively poor localisation and crowded field make it impossible to identify a single candidate based only on spatial coincidence. With the recent results of \citet{hurleywalker2024} and \citet{deruiter2024}, our understanding of the optical counterpart population of long-period radio transients is growing. A deeper understanding of the detectability and characteristics of this population will allow us to revisit \gleamx\ with renewed prior information to better quantify our ability to extract the counterpart from the data.

\section{Conclusion}
\label{sec:conclusion}

We have presented a comprehensive search for optical and NIR variability in the vicinity of the long-period radio transient \gleamx\ and found no convincing counterpart. We obtain the following conclusions.

\begin{itemize}
    \item The local stellar population, characterised by MUSE optical spectra, is dominated by low-mass K/M stars with no strong spectral peculiarities identified, although our sensitivity is not sufficient to probe the typical luminosities of isolated WDs.
    \item No source was found to exhibit periodicity or variability in either high-speed optical ULTRACAM data taken at 10\,second cadence over a period of 1.9\,hours, or lower-cadence NIR VVV survey epochs with revisit times of weeks over a period of $\sim3$\,years.
    \item Two Balmer emission line sources were identified, but their characteristics are adequately described by chromospheric activity in isolated M-dwarfs and are unlikely to indicate a close WD binary counterpart.
    \item No source exhibits a radial velocity curve strongly showing expectations for a close binary system with $P_\textrm{orb} \lesssim $\,hours, ruling out their presence to our limits (excluding the possibility of an almost-face-on system). What RV variation is seen for a few sources is likely due to other astrophysical considerations (e.g. rotation \citet{jenkins2009}) or under estimated uncertainties.
    \item We rule out the presence of a hot sub-dwarf in the localisation, which we would have expected to detect at high SNR.
    \item We determine our search is insensitive to an (effectively) isolated magnetar or WD, or compact binary system (WD/NS + WD/NS) counterpart, and so they remain viable scenarios.
    \item Most close binary systems RV amplitudes are within reach of MUSE even for highly inclined systems, but further monitoring is needed to provide robust evidence of any close-binary system(s) with interesting periods.
    \item Further observations of \gleamx\ during an active radio phase with high spatial resolution, which will allow for an unambiguous spatial association with a presented MUSE (or a hitherto undetected) source found during a radio-quiet period, may be the most likely route to unambiguously identifying its counterpart.
\end{itemize}

\section*{Acknowledgements}

We thank the anonymous reviewer for suggestions that improved the clarity of the manuscript.
JDL acknowledges support from a UK Research and Innovation Future Leaders Fellowship (MR/T020784/1). VSD and ULTRACAM are funded by the Science and Technology Facilities Council (grant ST/Z000033/1). SKA acknowledges funding from UKRI in the form of a Future Leaders Fellowship (grant no. MR/T022868/1, MR/Y034147/1). AAC acknowledges support through the European Space Agency (ESA) research fellowship programme. AJL has received funding from the European Research Council (ERC) under the European Union’s Seventh Framework Programme (FP7-2007-2013) (Grant agreement No. 725246). IP acknowledges support from a Royal Society University Research Fellowship (URF\textbackslash R1\textbackslash 231496). DTHS acknowledges support from the Science and Technology Facilities Council (STFC, grant numbers ST/T007184/1, ST/T003103/1, ST/T000406/1 and ST/Z000165/1)

This publication makes use of data products from the Two Micron All Sky Survey, which is a joint project of the University of Massachusetts and the Infrared Processing and Analysis Center/California Institute of Technology, funded by the National Aeronautics and Space Administration and the National Science Foundation.

This work made use of Astropy:\footnote{\url{http://www.astropy.org}} a community-developed core Python package and an ecosystem of tools and resources for astronomy \citep{astropy2013, astropy2018, astropy2022}. 

Based on observations collected at the European Organisation for Astronomical Research in the Southern Hemisphere under ESO programme(s) 108.23MN.

\section*{Data Availability}

Raw and processed data used are available via the European Southern Observatory data archives. Data used in the analysis or presented in figures are available upon reasonable request to the corresponding author.


\bibliographystyle{mnras}
\bibliography{references}



\appendix

\section{Optical sources in the vicinity of GLEAM-X\,J1627}
\label{app:specific_sources}

In \cref{tab:sourcesofinterest} we show the coordinates for those sources specifically addressed in the text or figures of this work, as well as all those detected within 4\,arcsec (2$\sigma$ localisation) of \gleamx. The sources were also photometered using a white-light image of the MUSE data cube. This was calibrated to \textit{Gaia} G-band using cross matches within the field of view. MUSE does not extend as blue as \textit{Gaia} G-band (cutting off at 470 and 400\,nm, respectively), although typically differences between the two magnitudes are $\lesssim0.1$\,mag. \textit{Gaia} cross-matches were found through a cone search of Data Release 3. The sources were photometered using {\sc sep} \citep{sep}; magnitudes are not presented for insignificant detections.

\begin{table*}
\begin{adjustbox}{width=\textwidth,totalheight=0.9\textheight,keepaspectratio}
\begin{tabular}{ccccccccccc}
\hline
id & \emph{Gaia} DR3 ID & RA & Decl. & M$_\textrm{MUSE}$ & $\sigma$(M$_\textrm{MUSE}$) & separation & ULTRACAM & VVV & RV & Emission \\
 &  & deg & deg & AB mag & AB mag & arcsecond &  &  &  &  \\
\hline
247 & 5933733753139994240 & 246.99647 & -52.58764 & 20.81 & 0.04 & 11.64 &  &  & \checkmark &  \\
254 & --- & 247.00011 & -52.58751 & --- & --- & 11.75 &  &  & \checkmark &  \\
267 & --- & 247.00052 & -52.58725 & 21.92 & 0.06 & 11.33 &  &  & \checkmark &  \\
268 & --- & 246.99664 & -52.58717 & --- & --- & 9.92 &  &  & \checkmark &  \\
273 & --- & 246.99539 & -52.58707 & 23.01 & 0.11 & 10.69 &  &  & \checkmark &  \\
278 & --- & 247.00093 & -52.58684 & --- & --- & 10.62 &  &  & \checkmark &  \\
279 & --- & 246.99452 & -52.58686 & 23.19 & 0.12 & 11.19 &  &  & \checkmark &  \\
282 & --- & 247.00017 & -52.58679 & --- & --- & 9.51 &  &  & \checkmark &  \\
285 & 5933733753138856832 & 246.99906 & -52.58672 & 20.39 & 0.03 & 8.29 & \checkmark & \checkmark & \checkmark &  \\
287 & 5933733753139670912 & 246.99639 & -52.58665 & 20.96 & 0.04 & 8.35 & \checkmark &  & \checkmark &  \\
291 & 5933733753136325504 & 247.00168 & -52.58666 & 19.92 & 0.03 & 11.27 & \checkmark & \checkmark & \checkmark &  \\
295 & --- & 246.99942 & -52.58657 & 21.46 & 0.05 & 8.05 &  &  & \checkmark &  \\
299 & --- & 246.99591 & -52.58651 & 23.17 & 0.12 & 8.37 &  &  & \checkmark &  \\
300 & --- & 246.99542 & -52.58649 & 22.92 & 0.10 & 8.94 &  &  & \checkmark &  \\
306 & --- & 246.99820 & -52.58627 & 21.11 & 0.04 & 6.30 &  & \checkmark & \checkmark &  \\
309 & --- & 247.00129 & -52.58624 & 22.71 & 0.09 & 9.62 &  &  & \checkmark &  \\
313 & 5933733753136323968 & 247.00254 & -52.58622 & 19.82 & 0.02 & 11.81 &  &  & \checkmark &  \\
314 & --- & 246.99694 & -52.58612 & --- & --- & 6.10 &  &  & \checkmark &  \\
318 & --- & 246.99795 & -52.58604 & --- & --- & 5.43 &  &  & \checkmark &  \\
319 & --- & 246.99636 & -52.58603 & 21.69 & 0.06 & 6.39 &  &  & \checkmark & \checkmark \\
320 & --- & 246.99508 & -52.58602 & 21.58 & 0.05 & 8.21 & \checkmark &  & \checkmark &  \\
324 & --- & 247.00072 & -52.58600 & 21.80 & 0.06 & 8.09 &  &  & \checkmark &  \\
335 & --- & 247.00155 & -52.58572 & 22.85 & 0.10 & 9.02 &  &  & \checkmark &  \\
336 & --- & 247.00091 & -52.58570 & --- & --- & 7.79 & \checkmark &  & \checkmark &  \\
339 & --- & 247.00282 & -52.58570 & 22.00 & 0.07 & 11.53 &  &  & \checkmark & \checkmark \\
340 & --- & 246.99598 & -52.58567 & 22.24 & 0.07 & 5.90 &  &  & \checkmark &  \\
344 & --- & 247.00247 & -52.58566 & 22.36 & 0.08 & 10.76 &  &  & \checkmark &  \\
345 & --- & 246.99747 & -52.58563 & --- & --- & 4.10 &  &  & \checkmark &  \\
348 & --- & 247.00165 & -52.58557 & 21.92 & 0.06 & 8.98 &  &  & \checkmark &  \\
349 & --- & 246.99828 & -52.58551 & 23.06 & 0.11 & 3.62 &  &  & \checkmark &  \\
350 & --- & 247.00215 & -52.58557 & 22.37 & 0.08 & 9.98 &  &  & \checkmark &  \\
351 & --- & 246.99328 & -52.58553 & --- & --- & 10.76 &  &  & \checkmark &  \\
354 & --- & 246.99667 & -52.58546 & 23.23 & 0.12 & 4.34 &  &  & \checkmark &  \\
359 & --- & 246.99865 & -52.58542 & 22.70 & 0.09 & 3.60 &  &  & \checkmark &  \\
360 & 5933733753136360448 & 246.99401 & -52.58539 & 19.69 & 0.02 & 9.09 & \checkmark & \checkmark & \checkmark &  \\
367 & 5933733753138853248 & 246.99826 & -52.58525 & 20.59 & 0.03 & 2.72 & \checkmark & \checkmark & \checkmark &  \\
370 & --- & 246.99464 & -52.58518 & 22.22 & 0.07 & 7.55 &  &  & \checkmark &  \\
384 & 5933733753169960448 & 247.00197 & -52.58494 & 19.01 & 0.02 & 9.00 & \checkmark & \checkmark & \checkmark &  \\
385 & --- & 247.00121 & -52.58493 & 23.13 & 0.11 & 7.34 &  &  & \checkmark &  \\
389 & --- & 247.00063 & -52.58484 & 23.34 & 0.13 & 6.03 &  &  & \checkmark &  \\
390 & --- & 246.99819 & -52.58484 & --- & --- & 1.28 &  &  & \checkmark &  \\
395 & --- & 246.99357 & -52.58481 & --- & --- & 9.57 &  &  & \checkmark &  \\
396 & --- & 246.99266 & -52.58478 & --- & --- & 11.53 &  &  & \checkmark &  \\
400 & 5933733753138309120 & 246.99949 & -52.58470 & 20.15 & 0.03 & 3.49 & \checkmark & \checkmark & \checkmark &  \\
412 & --- & 247.00241 & -52.58454 & 23.03 & 0.11 & 9.82 &  &  & \checkmark &  \\
415 & --- & 246.99704 & -52.58447 & --- & --- & 1.93 &  &  &  &  \\
424 & 5933733753136357760 & 246.99768 & -52.58430 & 20.16 & 0.03 & 0.96 & \checkmark & \checkmark & \checkmark &  \\
435 & 5933733748823423104 & 246.99627 & -52.58406 & 18.79 & 0.01 & 3.97 & \checkmark & \checkmark & \checkmark &  \\
439 & 5933733753136354432 & 246.99958 & -52.58400 & 19.43 & 0.02 & 4.10 & \checkmark & \checkmark & \checkmark &  \\
443 & --- & 246.99267 & -52.58395 & 20.28 & 0.03 & 11.66 &  &  & \checkmark &  \\
451 & --- & 247.00310 & -52.58382 & 23.23 & 0.12 & 11.61 &  &  & \checkmark &  \\
454 & --- & 246.99688 & -52.58374 & 23.47 & 0.14 & 3.63 &  &  &  &  \\
455 & --- & 246.99327 & -52.58376 & --- & --- & 10.53 &  &  & \checkmark &  \\
463 & --- & 246.99507 & -52.58362 & 22.04 & 0.07 & 7.04 &  &  & \checkmark &  \\
467 & --- & 246.99930 & -52.58356 & 21.84 & 0.06 & 4.63 &  &  & \checkmark &  \\
468 & --- & 246.99756 & -52.58356 & --- & --- & 3.58 &  &  & \checkmark &  \\
469 & --- & 246.99283 & -52.58359 & --- & --- & 11.63 &  &  & \checkmark &  \\
480 & --- & 247.00284 & -52.58340 & 23.12 & 0.11 & 11.50 &  &  & \checkmark &  \\
494 & --- & 247.00122 & -52.58325 & --- & --- & 8.58 &  &  & \checkmark &  \\
495 & --- & 246.99940 & -52.58328 & --- & --- & 5.55 &  &  & \checkmark &  \\
496 & --- & 246.99346 & -52.58323 & --- & --- & 10.80 &  &  & \checkmark &  \\
525 & 5933733753138848640 & 246.99752 & -52.58280 & 20.87 & 0.04 & 6.29 &  &  & \checkmark &  \\
526 & --- & 246.99875 & -52.58277 & 22.17 & 0.07 & 6.60 &  &  & \checkmark &  \\
529 & 5933733753138311168 & 246.99550 & -52.58267 & 20.44 & 0.03 & 8.52 & \checkmark & \checkmark & \checkmark &  \\
534 & --- & 246.99600 & -52.58256 & 22.64 & 0.09 & 8.23 &  &  & \checkmark &  \\
539 & --- & 247.00200 & -52.58253 & --- & --- & 11.47 &  &  & \checkmark &  \\
540 & --- & 246.99960 & -52.58252 & --- & --- & 8.10 &  &  & \checkmark &  \\
547 & --- & 247.00046 & -52.58241 & --- & --- & 9.45 &  &  & \checkmark &  \\
548 & 5933733753138848512 & 246.99783 & -52.58242 & 20.54 & 0.03 & 7.61 & \checkmark & \checkmark & \checkmark &  \\
557 & --- & 246.99569 & -52.58222 & 23.26 & 0.12 & 9.61 &  &  & \checkmark &  \\
563 & --- & 247.00115 & -52.58209 & 20.91 & 0.04 & 11.26 &  &  & \checkmark &  \\
564 & --- & 246.99449 & -52.58208 & 23.38 & 0.13 & 11.59 &  &  & \checkmark &  \\
572 & 5933733753176404864 & 246.99810 & -52.58195 & 20.09 & 0.03 & 9.31 &  & \checkmark & \checkmark &  \\
574 & --- & 246.99979 & -52.58195 & 21.24 & 0.05 & 10.16 &  &  & \checkmark &  \\
575 & --- & 246.99498 & -52.58192 & --- & --- & 11.37 &  &  & \checkmark &  \\
590 & 5933733753136366976 & 247.00056 & -52.58171 & 18.04 & 0.01 & 11.68 &  &  & \checkmark &  \\
591 & --- & 246.99749 & -52.58172 & 21.38 & 0.05 & 10.17 &  & \checkmark & \checkmark &  \\
592 & --- & 246.99631 & -52.58171 & --- & --- & 10.75 &  &  & \checkmark &  \\
598 & --- & 246.99941 & -52.58157 & 22.90 & 0.10 & 11.14 &  &  & \checkmark &  \\
599 & --- & 246.99710 & -52.58157 & 21.78 & 0.06 & 10.79 &  &  & \checkmark &  \\
\hline
\end{tabular}
\end{adjustbox}
\caption{Data for sources in the vicinity of \gleamx\ analysed during this work. All sources within 2$\sigma$ of the localisation or otherwise mentioned in the manuscript are included. M$_\textrm{MUSE}$ is the magnitude of the source in a white-light image of the MUSE data cube, calibrated to \textit{Gaia} G-band. `separation' is that of the source from \gleamx. `ULTRACAM` and `VVV` denote if the source was also detected and analysed in those data (see text).}
\label{tab:sourcesofinterest}
\end{table*}

\section{Spectra and Radial velocity curves of nearby sources}
\label{app:2sig}

\begin{figure*}
    \centering
    \includegraphics[width=\linewidth]{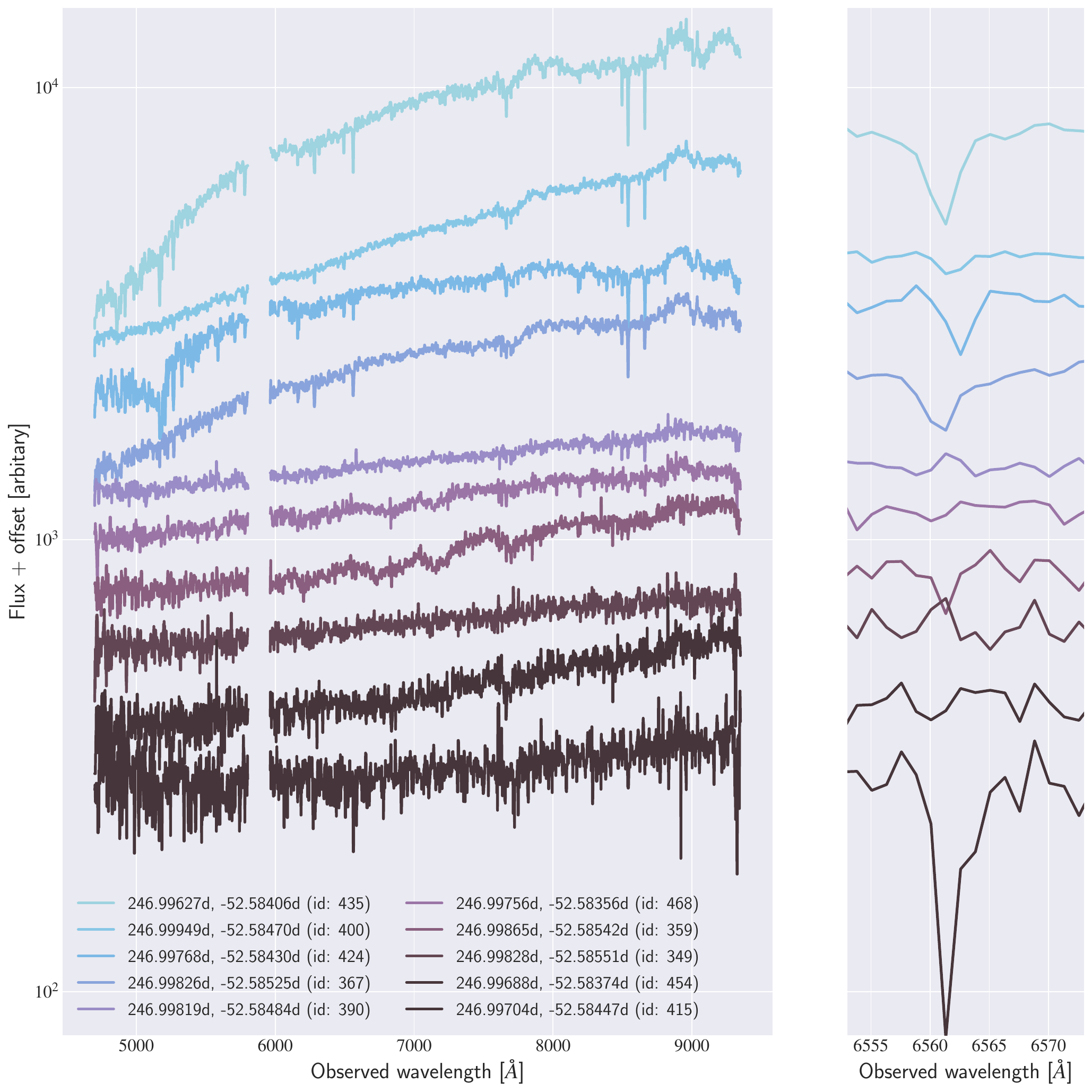}
    \caption{Extracted spectra of sources within the 2$\sigma$ localisation of \gleamx. The right panel offers a zoom-in around the location of H$\alpha$. The legend indicates their Right Ascension and Declination coordinates, with an accompanying source id matching those of \cref{tab:sourcesofinterest}.}
    \label{fig:spec2sig}
\end{figure*}

\begin{figure*}
    \centering
    \includegraphics[width=0.75\linewidth, clip, trim=0 50 0 0]{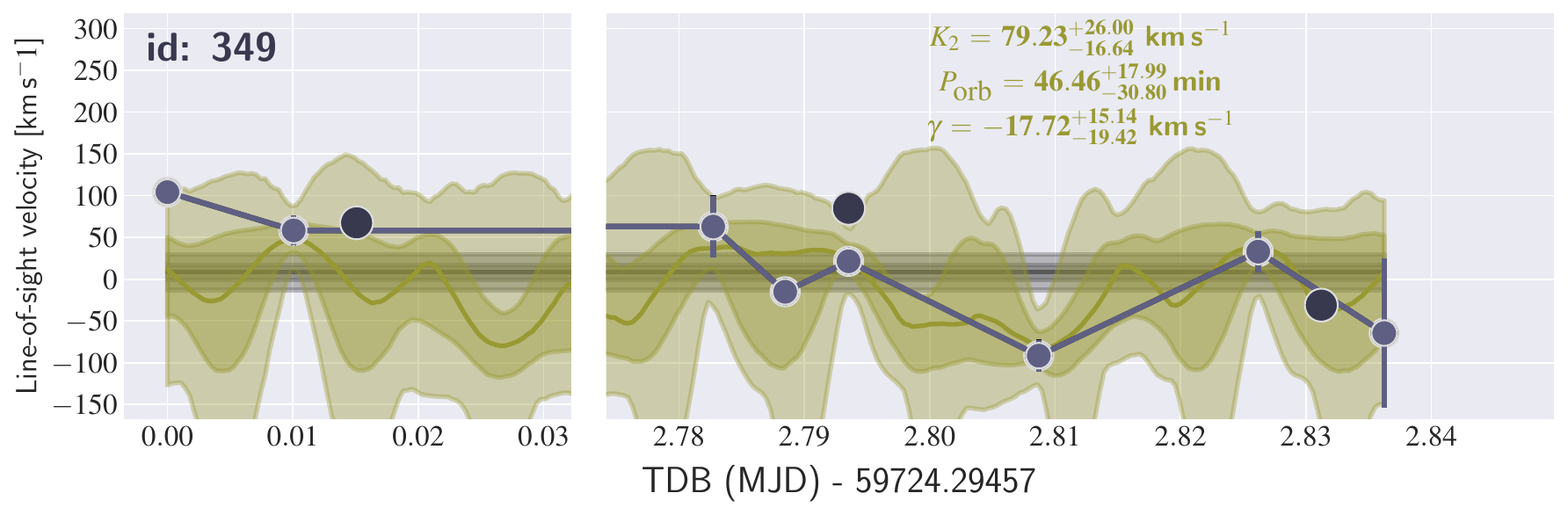}
    \includegraphics[width=0.75\linewidth, clip, trim=0 50 0 0]{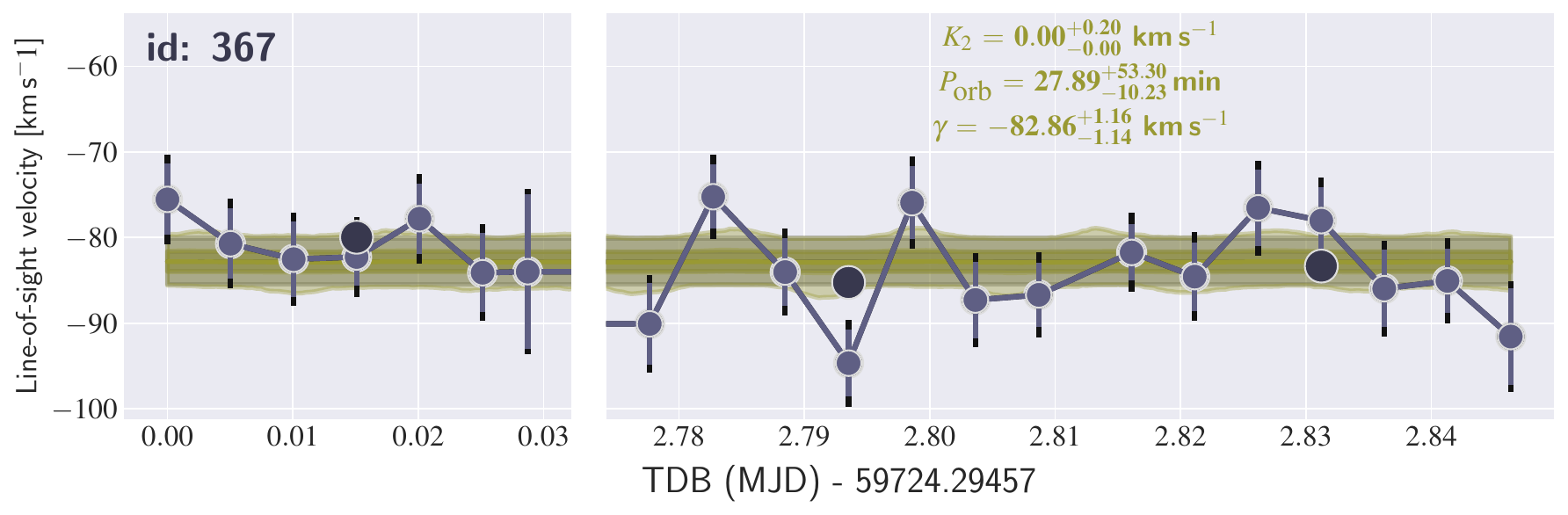}
    \includegraphics[width=0.75\linewidth, clip, trim=0 50 0 0]{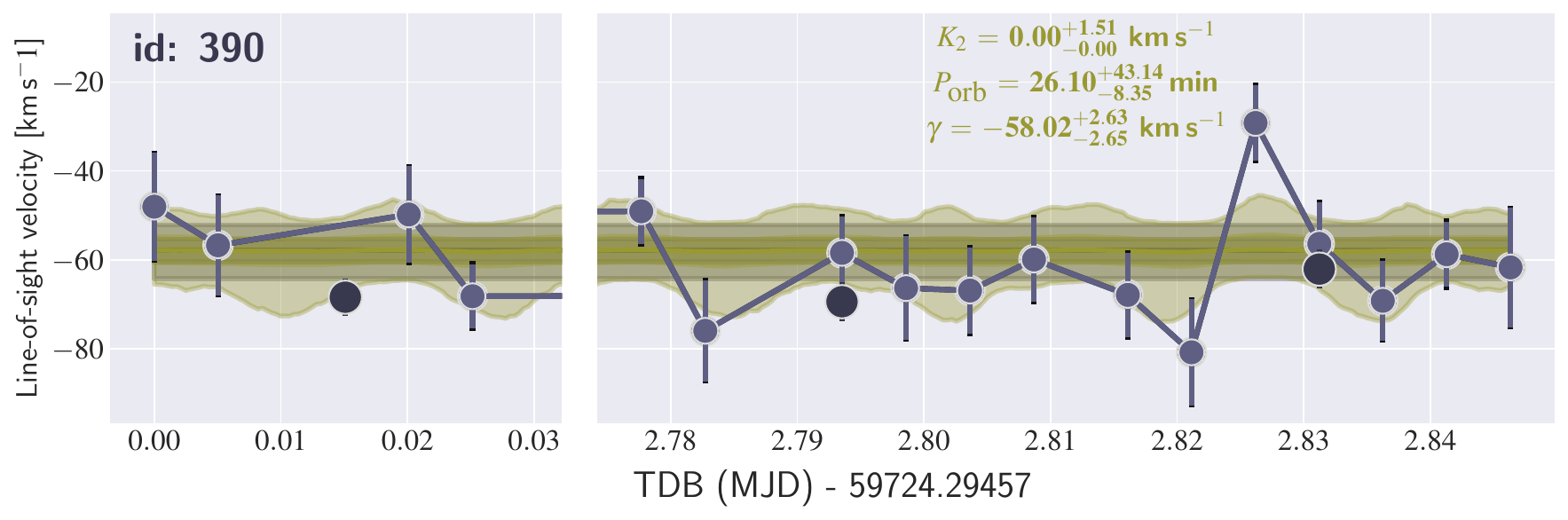}
    \includegraphics[width=0.75\linewidth, clip, trim=0 50 0 0]{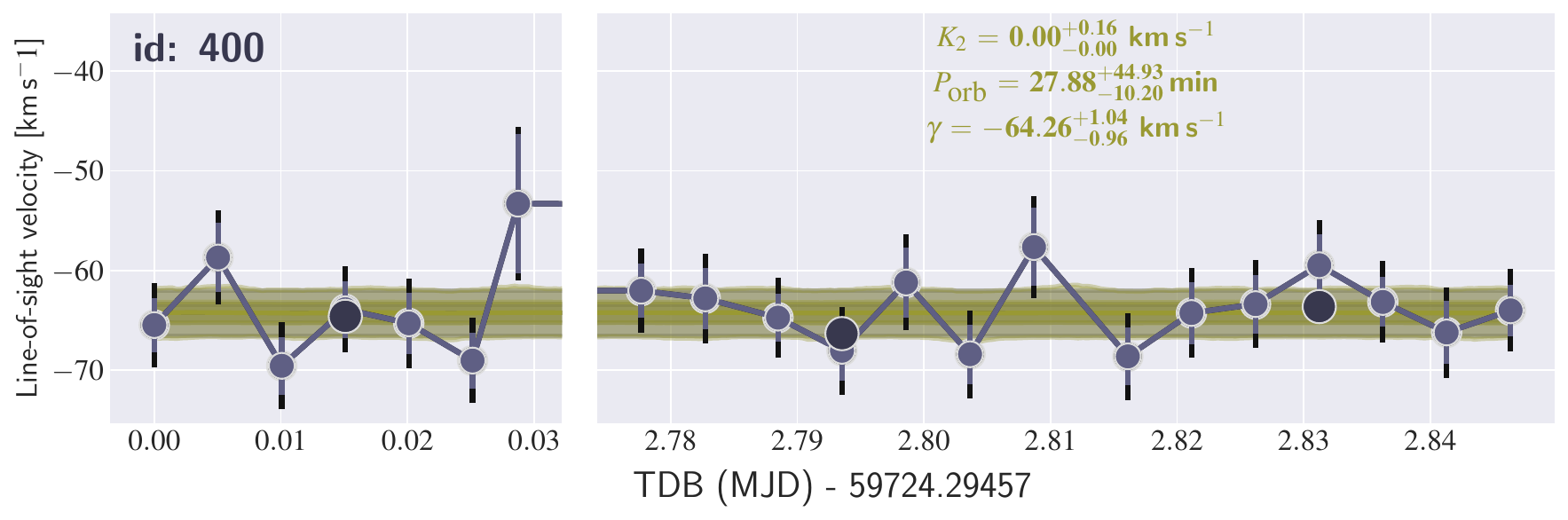}
    \includegraphics[width=0.75\linewidth, clip, trim=0 50 0 0]{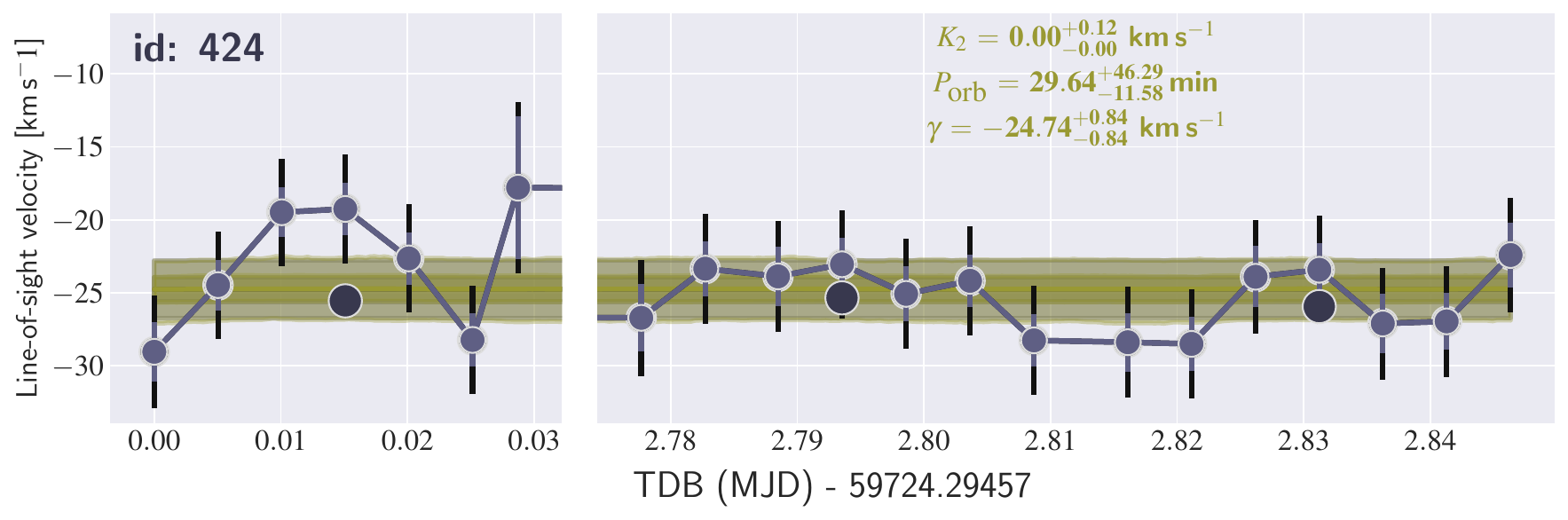}
    \includegraphics[width=0.75\linewidth]{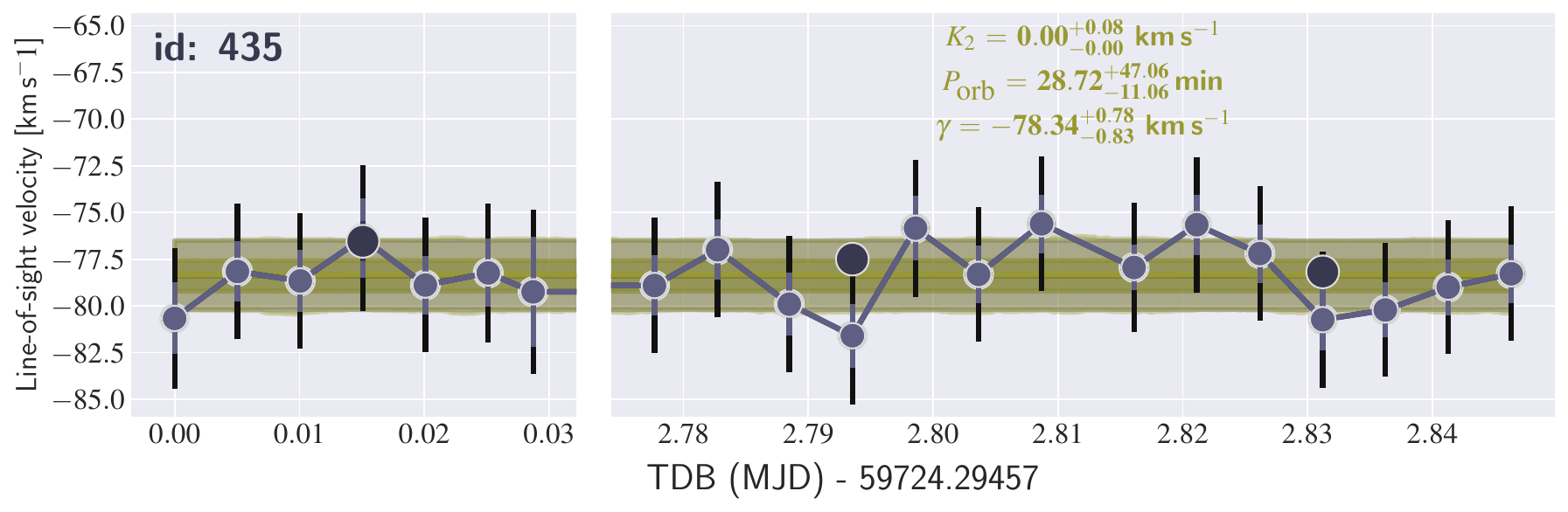}
    \caption{Same as \cref{fig:rvvar}, but shown for additional sources within the 2$\sigma$ localisation of \gleamx.}
    \label{fig:rv2sig}
\end{figure*}

In \cref{fig:spec2sig} we show extracted spectra for all detected sources within 4\,arcsec (2$\sigma$ localisation) of \gleamx. In \cref{fig:rv2sig} we show radial velocity curves of the subset of those sources where we were able to perform the analysis of \cref{sec:meth_rvvar} and determine a radial velocity curve. Not shown are source ids 454 and 415, for which we could not determine reliable RVs for enough individual epochs, and source ids 359 and 468, which are presented in \cref{fig:rvvar,fig:rvvar2}, respectively.

\section{Light curve of source id 548}
\label{app:lightcurve}

\begin{figure*}
    \centering
    \includegraphics[width=0.75\linewidth]{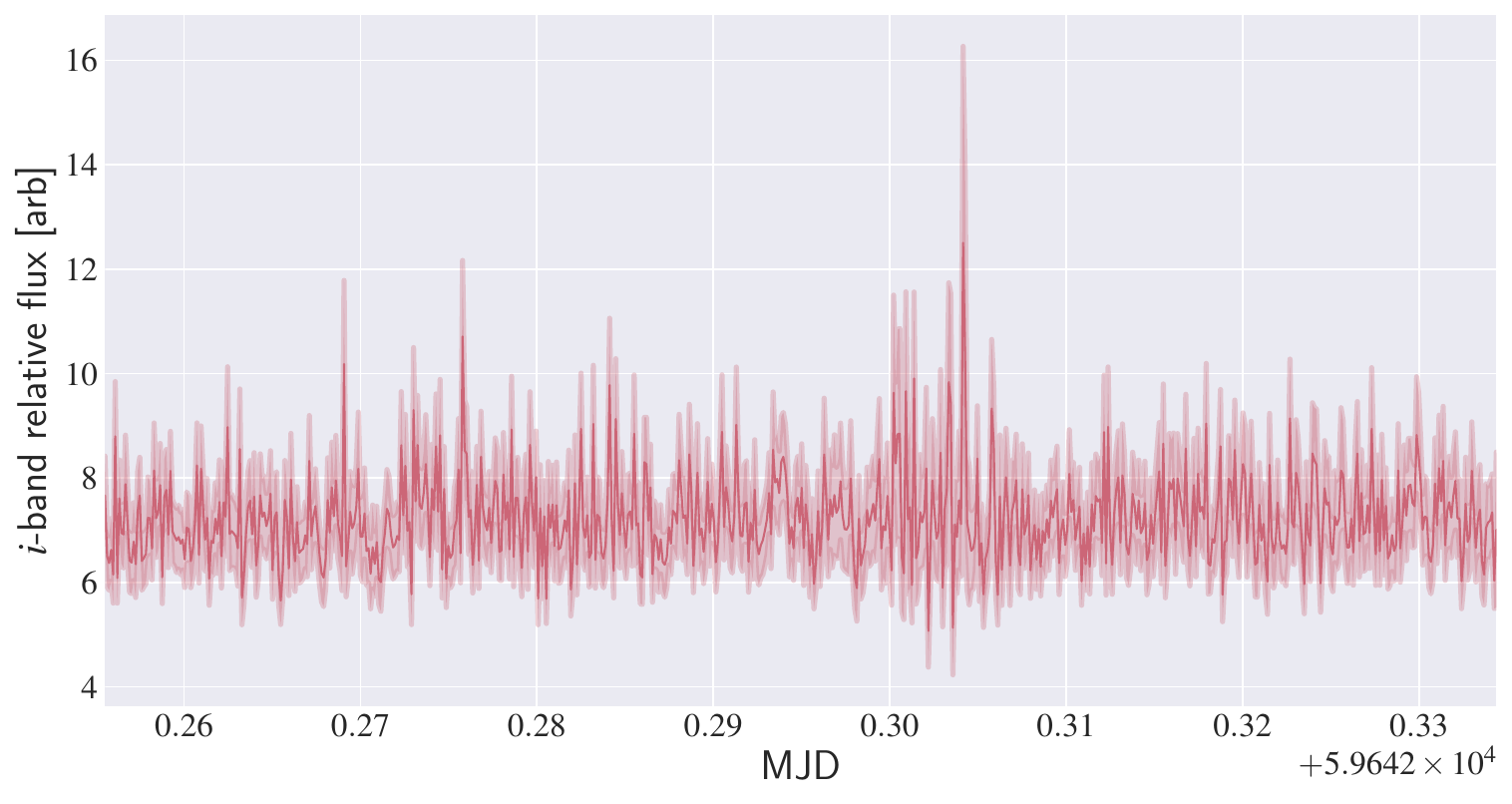}
    \caption{Light curve of aperture containing source id 548 from ULTRACAM. The increase in noise at MJD=59642.30-59642.35 is due to a deterioration of the atmospheric seeing. No periodic variability was found from analysis of this light curve (see text).}
    \label{fig:548uclightcurve}
\end{figure*}

\begin{figure*}
    \centering
    \includegraphics[width=0.75\linewidth]{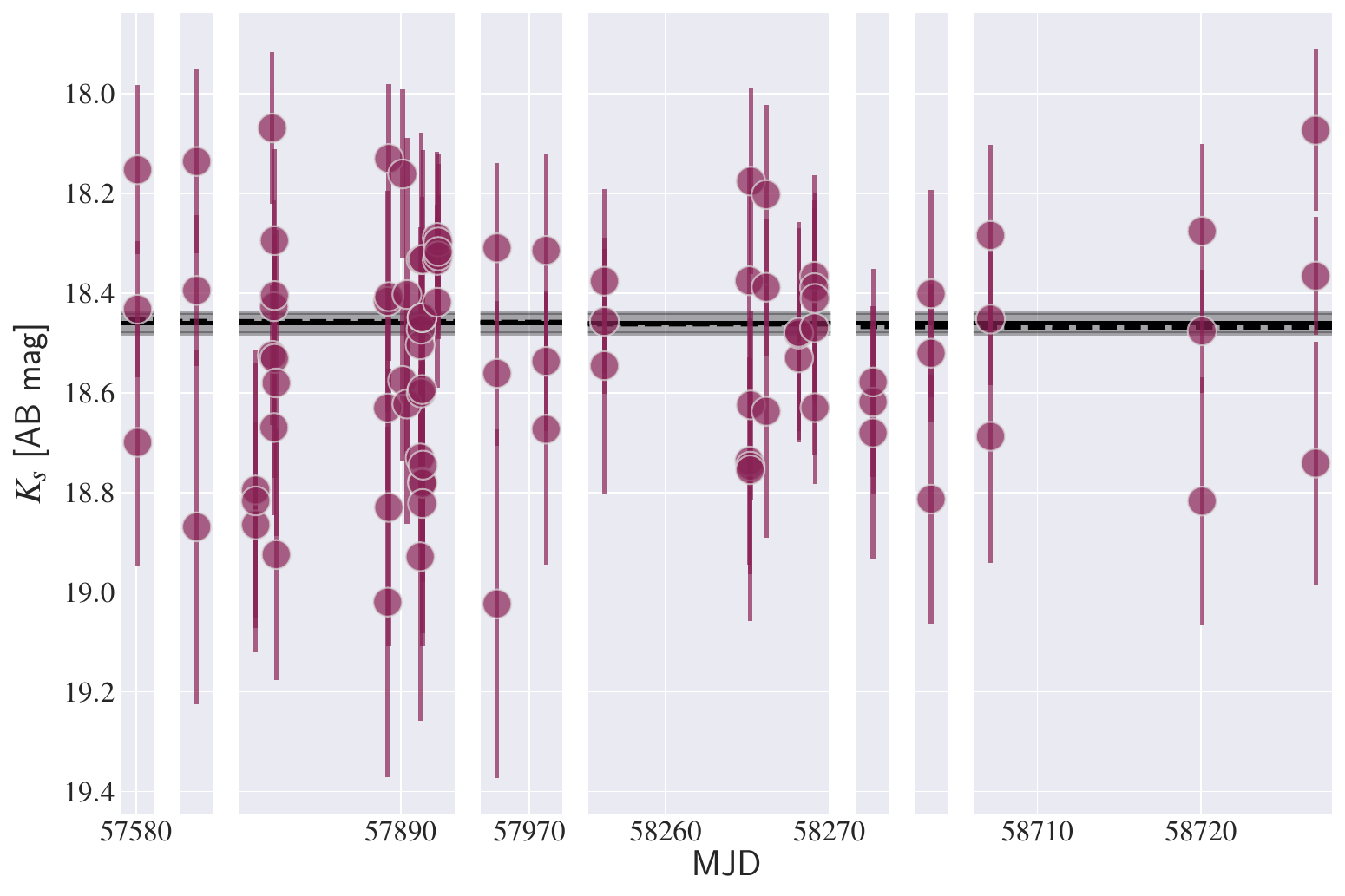}
    \caption{Light curve of source id 548 from VVV. The solid black line and grey shaded region show the model fit of a light curve as a constant flux model. The (almost indistinguishable) dashed black line indicates the best linear in time fit to the flux evolution. There is no evidence of outbursting or variability in individual photometric measurements, nor a long-term evolution of the flux level of the source.}
    \label{fig:548vvvlightcurve}
\end{figure*}

Source id 548 is the only source which passed our initial quality checks for RV variability and was bright enough to be detected in our photometric searches of ULTRACAM and VVV data. Its light curve is shown in \cref{fig:548uclightcurve,fig:548vvvlightcurve}. The source is significantly blended in ULTRACAM with source ids 525 and 572, and a conjoined aperture encompassing these three was used. The data display no indications of periodic or outbursting variability, in line with all other sources searched photometrically (see \cref{sec:res_photvar,fig:vvv_residuals}).

\bsp	
\label{lastpage}
\end{document}